\documentclass[sts]{imsart}

\RequirePackage{amsthm,amsmath,amsfonts,amssymb}
\RequirePackage[authoryear]{natbib}
\RequirePackage[colorlinks,citecolor=blue,urlcolor=blue]{hyperref}
\RequirePackage{graphicx}

\usepackage{amsmath}
\usepackage{amsfonts}
\usepackage{amssymb}
\usepackage{amsthm}
\usepackage{color}
\usepackage{subcaption}
\usepackage[dvipsnames,svgnames,x11names,hyperref]{xcolor}
\usepackage{pifont}

\usepackage{dsfont}
\hypersetup{linkcolor=RoyalBlue,citecolor=RoyalBlue,urlcolor=RoyalBlue}
\usepackage{tikz}
\usepackage{caption}
\usepackage{float}
\usetikzlibrary{arrows.meta}
\usetikzlibrary{calc}

\usepackage{comment}
\def\rmd{\mathrm{d}}

\def\rmd{\mathrm{d}}
\floatstyle{ruled}
\usepackage{algorithm, algorithmic}

\startlocaldefs
\theoremstyle{plain}

\theoremstyle{remark}

\endlocaldefs

\begin{document}

\begin{frontmatter}
\title{Diffusion Schr\"{o}dinger Bridges for Bayesian Computation}
\runtitle{Diffusion Schr\"{o}dinger Bridges for Bayesian Computation}

\begin{aug}
\author[A]{\fnms{Jeremy}~\snm{Heng}\ead[label=e1]{heng@essec.edu}},
\author[B]{\fnms{Valentin}~\snm{De Bortoli}\ead[label=e2]{valentin.debortoli@gmail.com}}
\and
\author[C]{\fnms{Arnaud}~\snm{Doucet}\ead[label=e3]{doucet@stats.ox.ac.uk}}


\address[A]{Jeremy Heng is Assistant Professor, ESSEC Business School, Singapore 139408, Singapore\printead[presep={\ }]{e1}.}

\address[B]{Valentin De Bortoli is Research Scientist, Center for Sciences of Data, ENS Ulm, Paris, France\printead[presep={\ }]{e2}.}

\address[C]{Arnaud Doucet is Professor, Department of Statistics, University of Oxford, Oxford OX1 3LB, U.K.\printead[presep={\ }]{e3}.}

\end{aug}

\begin{abstract}
Denoising diffusion models are a novel class of generative models that have recently become extremely popular in machine learning. In this paper, we describe how such ideas can also be used to sample from posterior distributions and, more generally, any target distribution whose density is known up to a normalizing constant. The key idea is to consider a forward ``noising'' diffusion initialized at the target distribution which ``transports'' this latter to a normal distribution for long diffusion times. The time-reversal of this process, the ``denoising'' diffusion, thus ``transports'' the normal distribution to the target distribution and can be approximated so as to sample from the target. To accelerate simulation, we show how one can introduce and approximate a Schr\"{o}dinger bridge between these two distributions, i.e. a diffusion which transports the normal to the target in finite time. 
\end{abstract}


\end{frontmatter}

\section{Introduction}
The use of diffusion processes as a mathematical model is ubiquitous in many scientific disciplines. In differential form, such a process $(X_t)_{t\in[0,T]}$ in $\mathbb{R}^d$ is defined by a stochastic differential equation (SDE) (see e.g. \citet{oksendal2003stochastic} and \citet{klebaner2012introduction} for textbooks on the subject) 
\begin{align}
    \rmd X_t = f(t,X_t)\rmd t + \sigma(t,X_t)\rmd B_t.
\end{align}
The above describes infinitesimal changes in the process as a sum of deterministic changes driven by a drift function $f:[0,T]\times\mathbb{R}^d\rightarrow\mathbb{R}^d$ for an infinitesimal time step, and random fluctuations given by a diffusion matrix $\sigma:[0,T]\times\mathbb{R}^d\rightarrow\mathbb{R}^{d\times d}$ and infinitesimal changes of a standard Brownian motion in $\mathbb{R}^d$. 
The interface between statistics and diffusion models have involved two main threads: statistical inference for these models and their use for Monte Carlo simulation.
In the first, one considers inference for the parameters of $f$ and $\sigma$ when observations of the process are collected at various time points. This has generated a large and comprehensive literature with frequentist and Bayesian estimators, under various types of observation regimes, and at various frequencies (see textbooks by \citet{iacus2008simulation} and \citet{kessler2012statistical} and references therein). In the second, following ideas from statistical mechanics, the use of certain diffusion processes and their long time behaviour have been adopted to simulate from complex and high-dimensional distributions in computational statistics. For example, by simulating Langevin dynamics whose equilibrium distribution is the target distribution of interest, one progressively transforms an easy-to-sample reference distribution to the desired target distribution. Indeed the time-discretization of such Langevin dynamics have lead to efficient gradient-based Markov chain Monte Carlo (MCMC) algorithms for sampling problems in Bayesian statistics \citep{grenander1994representations,roberts1996exponential}. 

This article is based on denoising diffusions, which provide an alternative approach to transform a simple reference distribution to a target distribution. These diffusions were recently introduced in the machine learning literature to obtain generative models \citep{Sohl-DicksteinW15,ho2020denoising,song2020score}. In contrast to running a single realization of Langevin dynamics to obtain a large number of correlated samples for MCMC, a single run of a denoising diffusion is used to simulate one independent sample from the generative model.
Given access to a large dataset, say of images or proteins, generative models produce new synthetic data with approximately the same distribution. Many generative modeling techniques have been proposed in the literature, but the introduction of denoising diffusions has recently revolutionized this field as these models are easier to train than alternatives, and they provide state-of-the-art performance in many domains. 
The main idea of these techniques is to progressively transform the empirical data distribution into a normal distribution through a forward ``noising'' diffusion process. By simulating an approximation of the time-reversal of this process, a ``denoising'' diffusion, we can generate new samples that closely follow the data distribution.

After introducing notation in Section \ref{sec:notation}, we review in Section \ref{sec:DDPS} how one can extend these ideas to perform approximate posterior simulation in scenarios where one can only sample from the prior and simulate synthetic observations of the likelihood \citep{song2020score}. A limitation of this approach is that it requires running the forward noising diffusion long enough to ``transport'' the posterior distribution into an approximately normal distribution. Hence the corresponding approximate time-reversal also needs to be run for a long time to sample from the posterior. In Section \ref{sec:DSB-PS}, we show how a Schr\"odinger bridge formulation of this problem can be exploited to transport in finite time the posterior into a normal and vice-versa \citep{shi2022conditional}, hence accelerating posterior simulation. 

In scenarios where one only has access to an unnormalized density of a target distribution, which might not correspond to the posterior distribution of a Bayesian model, we show in Section \ref{sec:DDGS} that the ideas behind denoising diffusion models remain applicable. In this setting, one has to optimize an alternative criterion to approximate the time-reversal of the forward noising diffusion \citep{vargasDDSampler2023}. Finally we present a Schr\"odinger bridge extension of this algorithm to speed up simulation in Section \ref{sec:general_SB}. To aid the reading of this article, the features of algorithms covered are summarized in Table \ref{tab:summary}.
We conclude with some discussions in Section \ref{sec:discuss}.

\section{Notation}\label{sec:notation}

We will use throughout this paper the following notation. We will write $x\in\mathcal{X}$ to denote latent states or parameters of a statistical model, and $y\in\mathcal{Y}$ to denote observations. 
For example in Euclidean spaces, we have $\mathcal{X}=\mathbb{R}^d$ and $\mathcal{Y}=\mathbb{R}^p$ for $p,d\in\mathbb{N}$. The Euclidean norm of $x\in\mathbb{R}^d$ is denoted by $\|x\|$. 
For $f:\mathbb{R}^d\rightarrow\mathbb{R}$, we denote its gradient with respect to $x$ as $\nabla_xf(x)$. 
We write $\mathcal{N}(\mu,\Sigma)$ to denote the normal distribution with mean $\mu$ and covariance $\Sigma$, and $x\mapsto\mathcal{N}(x;\mu,\Sigma)$ for its probability density function. 
Given two probability measures $\mu$ and $\nu$ defined on a measurable space $\mathcal{Z}$, the Kullback--Leibler (KL) divergence from $\mu$ to $\nu$ is defined as
$\mathrm{KL}(\mu|\nu) = \int_{\mathcal{Z}} \log(\rmd \mu / \rmd \nu)(z) \rmd \mu(z)$ if $\mu \ll \nu$ and $+\infty$ otherwise. 
We denote the set of integers $[K] = \{1, \dots, K\}$. 
A path measure on $\mathcal{Z}$ is a probability
measure on $C([0,T], \mathcal{Z})$, where $C([0,T], \mathcal{Z})$ is the space of continuous functions from $[0,T]$ to $\mathcal{Z}$. If $\mathbb{P}$ is a path measure then we denote its \emph{time-reversal} as $\mathbb{P}^R$, defined such that if
$(X_t)_{t \in [0,T]} \sim \mathbb{P}$ then $(X_{T-t})_{t \in [0,T]} \sim \mathbb{P}^R$. We will also denote the time $t\in[0,T]$ marginal distribution of $\mathbb{P}$ by $\mathbb{P}_t$, and write its probability density function as $p_t$. 
Given a probability measure $\mu$ on $\mathcal{Z}_1$, 
and a Markov kernel $\mathrm{K}$ from $\mathcal{Z}_1$ to $\mathcal{Z}_2$, we define the probability measure 
$\mu \otimes \mathrm{K}$ on $\mathcal{Z}_1 \times \mathcal{Z}_2$ as $(\mu \otimes \mathrm{K})(\mathsf{A}_1 \times \mathsf{A}_2) =
\int_{\mathsf{A}_1} \rmd \mu(z_1) \mathrm{K}(z_1, \mathsf{A}_2)$ for any Borel set $\mathsf{A}_1 \times \mathsf{A}_2$. 
Finally, for notational ease, we will use $(X_t)_{t \in [0,T]}$ to denote \emph{forward} diffusion processes driven by Brownian motion $(B_t)_{t \in [0,T]}$, and $(Z_t)_{t \in [0,T]}$ to denote \emph{backward} diffusion processes driven by Brownian motion $(W_t)_{t \in [0,T]}$.

\begin{table*}[t]
  \centering
  \begin{tabular}{l||c|c|c|c|}
    Method name & Require samples & Require unnormalized density & Iterative & Section \\
    \hline
    Denoising Diffusion for Posterior Sampling (DDPS) & \ding{51}  & \ding{55} & No & Section \ref{sec:DDPS} \\
    Diffusion Schr\"{o}dinger Bridge for Posterior Sampling (DSB-PS)  & \ding{51} & \ding{55} & Yes & Section \ref{sec:DSB-PS} \\
    Denoising Diffusion for General Sampling (DDGS) & \ding{55} & \ding{51} & No & Section \ref{sec:DDGS} \\ 
    Diffusion Schr\"{o}dinger Bridge for General Sampling (DSB-GS) & \ding{55} & \ding{51} & Yes & Section \ref{sec:general_SB} \\
    \hline
  \end{tabular}
  \caption{Summary of algorithms in this article.}
  \label{tab:summary}          
\end{table*}


\section{Denoising diffusions for posterior sampling\label{sec:DDPS}}
We recall that $x\in\mathcal{X}$ denote latent states or parameters of a statistical model and $y\in\mathcal{Y}$ denote observations. Given a prior distribution $\mu(x)$ and likelihood function $x\mapsto g(y|x)$, Bayesian inference is based on the posterior distribution 
\begin{align}\label{eqn:posterior}
    p(x|y) = \frac{p(x,y)}{p(y)}, \quad p(x,y)=\mu(x)g(y|x),
\end{align}
where $p(y)=\int_{\mathcal{X}} \mu(x)g(y|x)\rmd x$ is the marginal
likelihood. We begin by describing an amortized variational inference procedure
that allows us to sample approximately from $p(x|y)$ for all possible
observations $y$. This method known as \emph{denoising diffusions} was
introduced by \citet{song2020score} for $\mathcal{X}=\mathbb{R}^d$. We will
restrict ourselves here to this setup but it has been extended to
general spaces and discrete spaces
\citep{austin2021structured,hoogeboom2021argmax,campbell2022continuous,benton2022denoising,de2022riemannian,huang2022riemannian}. In this Section and Section \ref{sec:DSB-PS}, we assume that we can obtain samples from the joint distribution $p(x,y)$. Practically we sample first $X \sim \mu(x)$ followed by $Y \sim g(y|X)$. 

We first define a forward ``noising" process $(X_t)_{t\in[0,T]}$ according to the Ornstein--Uhlenbeck (OU) process
\begin{align}\label{eqn:noising_SDE}
    \rmd X_t = -\tfrac{1}{2}X_t\rmd t + \rmd B_t, \quad X_0 \sim p(x|y),
\end{align}
where $(B_t)_{t\in[0,T]}$ is a standard $d$-dimensional Brownian motion. Writing the transition density of \eqref{eqn:noising_SDE} as $p_{t|0}(x_t|x_0)$, the distribution of $X_t$ is given by $p_t(x_t|y)=\int_{\mathcal{X}} p_{t|0}(x_t|x_0)p(x_0|y)\rmd x_0$. 
Since Equation \eqref{eqn:noising_SDE} can be seen as the SDE of a Langevin diffusion with the standard normal distribution $\mathcal{N}(0,I)$ as its stationary distribution, by ergodicity properties of the process, 
$p_T(x_T|y)$ for large time $T>0$ approaches $\mathcal{N}(0,I)$ for all $y$. The corresponding time-reversed process $(Z_t)_{t\in[0,T]} = (X_{T-t})_{t\in[0,T]}$ can be shown to satisfy (weakly) the SDE \citep{anderson1982reverse,haussmann1986time}
\begin{align}\label{eqn:denoising_SDE}
    \rmd Z_t = \tfrac{1}{2}Z_t \rmd t+\nabla_{z_t}\log p_{T-t}(Z_t|y)\rmd t + \rmd W_t,
\end{align}
with another standard $d$-dimensional Brownian motion $(W_t)_{t\in[0,T]}$ and
$Z_0\sim p_T(x_T|y)$. If one could simulate the backward ``denoising" process
$(Z_t)_{t\in[0,T]}$, then $Z_T$ would be a sample from the posterior
distribution $p(x|y)$.  Practically, we cannot sample from
(\ref{eqn:denoising_SDE}), so consider instead a diffusion approximating it of
the form
\begin{equation}\label{eqn:KL_proposalAVI}
    \rmd Z_t =  \tfrac{1}{2}Z_t \rmd t+s^{\theta}_{T-t}(Z_t,y)\rmd t + \rmd W_t,
  \end{equation}
  with $Z_0\sim \mathcal{N}(0,I)$.  For the diffusion (\ref{eqn:KL_proposalAVI})
  to approximate (\ref{eqn:denoising_SDE}), we need 
  i) the diffusion time $T$ to be large enough for $p_T(x_T|y)\approx\mathcal{N}(x_T;0,I)$;
  ii) an approximation of the
  Stein score function $s^{\theta}_t(x_t,y)\approx \nabla_{x_t}\log p_t(x_t|y)$
  for all $(t,x_t,y)\in[0,T]\times\mathcal{X}\times\mathcal{Y}$. We show next
  how such an approximation can be obtained.

Let $\mathbb{P}_y$ denote the path measure on $C([0,T],\mathcal{X})$ induced by \eqref{eqn:noising_SDE} with observations $y\in\mathcal{Y}$ and let $\mathbb{Q}^{\theta}_y$ be the law on $C([0,T],\mathcal{X})$ induced by the time-reversal of \eqref{eqn:KL_proposalAVI}. We consider a parametric function class $\{s_t^{\theta}(x_t,y): \theta\in \Theta\}$ such as neural networks. We obtain $\theta$ approximating the score by minimizing the expected KL divergence between $\mathbb{P}_y$ and $\mathbb{Q}^{\theta}_y$ over $y \sim p(y)$, which satisfies \citep[Theorem 1]{song2021maximum}
\begin{align}  
\mathcal{L}(\theta)&=2\mathbb{E}_{y \sim p(y)}[\mathrm{KL}(\mathbb{P}_y|\mathbb{Q}^{\theta}_y)] \nonumber
                       \\ & \textstyle \equiv \int_0^T\mathbb{E}_{\overline{p}}\left[\|s_t^{\theta}(X_t,Y) - \nabla_{x_t}\log p_{t|0}(X_t|X_0) \|^2\right]\rmd t,  \label{eqn:DSM_loss}
\end{align}
where `$\equiv$' means equality up to an additive constant and the second
expectation in (\ref{eqn:DSM_loss}) is w.r.t.
$\overline{p}(x_0,x_t,y) = p(x_0,y)p_{t|0}(x_t|x_0)$. This loss corresponds to a
\emph{denoising score matching} (DSM) loss \citep{vincent2011connection}. Under
the transition density of the OU process in \eqref{eqn:noising_SDE}, the
gradient appearing in \eqref{eqn:DSM_loss} is
$\nabla_{x_t}\log p_{t|0}(x_t|x_0)=\{x_0\exp(-t/2)-x_t\}/\{1-\exp(-t)\}$. The
loss $\mathcal{L}(\theta)$ can be minimized using stochastic gradient based
algorithms such as Adam \citep{kingma2014adam} if one can simulate
$(X_0,Y)\sim p(x,y)$ from the model \eqref{eqn:posterior}. Hence this approach
is applicable to problems where both the prior and the likelihood are
intractable but one can simulate parameters and synthetic data from them. From
this perspective, it is an alternative to Approximate Bayesian Computation (ABC)
methods
\citep{marin2012approximate,sisson2018handbook,beaumont2019approximate}. Empirical
comparisons between denoising diffusions, MCMC, and
ABC methods can be found in
\citep{benton2022denoising,sharrock2022sequential,shi2022conditional}.

Having found a minimizer $\theta$ of \eqref{eqn:DSM_loss}, the resulting
denoising diffusion posterior sampler (DDPS) described in Algorithm
\ref{alg:DDPS} involves simulating (\ref{eqn:KL_proposalAVI}) using a suitable
numerical integrator (see e.g. \citet{karraselucidating}) and returning $Z_T$ as
an approximate sample from the posterior distribution $p(x|y)$ in
\eqref{eqn:posterior}.  Various techniques that have been proposed to accelerate
denoising diffusion models are also applicable to DDPS
\citep{dockhorn2021score,salimans2022progressive}. Finally, we note that one can tailor the above
procedure to specific observations $y$ in the case where the likelihood function can be evaluated. This relies on the following alternative decomposition to estimate the score 
\begin{equation}\label{eq:decomposcore}
\nabla_{x_t} \log p_t(x_t|y)=\nabla_{x_t} \log \mu_t(x_t)+\nabla_{x_t} \log g_t(y|x_t)
\end{equation}
In this setting, we have that
$\mu_t(x_t)=\int_{\mathcal{X}} \mu(x_0) p_{t|0}(x_t|x_0)\rmd x_0$ is the
diffused prior and
$g_t(y|x_t)=\int_{\mathcal{X}} g(y|x_0) p_{0|t}(x_0|x_t)\rmd x_0$ is the modified
likelihood. The first term on the r.h.s. of \eqref{eq:decomposcore} can be
estimated by $s^\theta_t(x_t)$ with $\theta$ obtained by minimizing
$\textup{KL}(\mathbb{P}|\mathbb{Q}^{\theta})$, where $\mathbb{P}$ is induced by
the noising diffusion (\ref{eqn:noising_SDE}) initialized using
$X_0 \sim \mu(x)$ and $\mathbb{Q}^{\theta}$ is induced by the reversal of the
form (\ref{eqn:KL_proposalAVI}) with $s^\theta_t(x_t,y)$ replaced by
$s^\theta_t(x_t)$. 
Using evaluations of $g(y|x)$, the term $g_t(y|x_t)$ can be approximated using regression, \emph{conditional guidance} techniques \citep{chung2209diffusion}, or Monte Carlo methods \citep{pmlr-v202-song23k}. 

The next section presents a principled framework based on Schr\"{o}dinger bridges to accelerate training and sampling in DDPS.

\begin{algorithm}
  \protect\caption{Denoising diffusion for posterior sampling \label{alg:DDPS}}
  \begin{algorithmic}
\small
\STATE{\textbf{--- Training procedure:}}
\WHILE{not converged}
\STATE{Sample $(X_t^k)_{t \in [0,T]}$ using the SDE \eqref{eqn:noising_SDE} where $(X_0^k, Y^k) \sim p(x,y)$ for $k \in [K]$.}
\STATE{Approximate the loss \eqref{eqn:DSM_loss} using $((X_t^k)_{t \in [0,T]},Y^k)_{k\in[K]}$.}
\STATE{Update $\theta$ in $s_t^\theta$ using Adam optimizer.}
\ENDWHILE
\STATE{\textbf{--- Sampling procedure:}}
\STATE{Sample $(Z_t)_{t\in[0,T]}$ using the SDE \eqref{eqn:KL_proposalAVI} for the  observation $y$.}
\STATE{\textbf{return:} $Z_T$ approximately distributed according to $p(x|y)$.}
  \end{algorithmic}
\end{algorithm}

\section{Diffusion Schr\"{o}dinger bridges for posterior sampling\label{sec:DSB-PS}}
For DDPS to perform well, the diffusion time $T$ has to be long enough so that $p_T(x_T|y)\approx\mathcal{N}(x_T;0,I)$. 
\citet{shi2022conditional} considered the \emph{Schr\"{o}dinger bridge} (SB) formulation described in \citep{de2021diffusion,vargas2021solving,chenlikelihood} to improve DDPS in this setting. Let $\mathbb{D}$ denote the path measure on $C([0,T],\mathcal{Y})$ induced by 
\begin{align}\label{eqn:constant_SDE}
    \rmd Y_t = 0, \quad Y_0\sim p(y),
\end{align}
i.e. $Y_t=Y_0$ for all $t\in[0,T]$. The SB is defined by the following constrained KL minimization problem over probability measures on the path space $C([0,T],\mathcal{X}\times\mathcal{Y})$:
\begin{align}\label{eqn:SB_dynamic}
&\textstyle
  \Pi^{\star}=\arg\min_{\Pi}\{\mathrm{KL}(\Pi|\mathbb{P}): \Pi_0(x_0,y_0)=p(x_0,y_0), \nonumber \\
  & \textstyle  \qquad \qquad \quad  \quad \Pi_T(x_T,y_T)=\mathcal{N}(x_T;0,I)p(y_T)\}, 
\end{align}
where $\mathbb{P}=\mathbb{P}_{y_0}\otimes\mathbb{D}$ is the joint measure associated to \eqref{eqn:noising_SDE} and \eqref{eqn:constant_SDE}, and $\Pi_t$ denotes the time $t$ marginal distribution under $\Pi$. As the minimizer has the form $\Pi^{\star}=\mathbb{P}_{y_0}^{\star}\otimes\mathbb{D}$, by simulating a backward process $(Z_t)_{t\in[0,T]}$ with law $\mathbb{P}_y^{\star}$ and $Z_0$ initialized from $\Pi_T^{\star}(x_T|y)=\mathcal{N}(x_T;0,I)$, the terminal state $Z_T$ returns a sample from the posterior distribution $\Pi_0^{\star}(x_0|y)=p(x_0|y)$, $p(y)$-almost surely. 

To solve for the SB, we apply the \emph{iterative proportional fitting} (IPF) procedure \citep{fortet1940resolution,deming1940least,kullback1968probability}. This is defined by the following recursions for $n\in\mathbb{N}_0$:
\begin{align}
  &\textstyle\Pi^{2n+1}=\arg\min_{\Pi}\{\mathrm{KL}(\Pi|\Pi^{2n}): \nonumber \\
  & 
  \qquad \Pi_T(x_T,y_T)=\mathcal{N}(x_T;0,I)p(y_T)\}, \nonumber \\
    &\textstyle\Pi^{2n+2}=\arg\min_{\Pi}\{\mathrm{KL}(\Pi|\Pi^{2n+1}
      ): \nonumber \\ 
  & \qquad \qquad \Pi_0(x_0,y_0)=p(x_0,y_0)\}, 
\end{align}
initialized at $\Pi^0=\mathbb{P}$, which involves iterative KL-projections to impose the marginal distribution $\Pi_0(x_0,y_0)=p(x_0,y_0)$ at time $0$, and $\Pi_T(x_T,y_T)=\mathcal{N}(x_T;0,I)p(y_T)$ at time $T$. Convergence of the iterates $(\Pi^n)_{n\in\mathbb{N}}$ to the SB $\Pi^{\star}$ has been studied under various sets of assumptions \citep{ruschendorf1995convergence,chen2016entropic,de2021diffusion,leger2021gradient}. It can be shown that $\Pi^{2n+1}=\mathbb{P}_{y_T}^{2n+1}\otimes\mathbb{D}$ and $\Pi^{2n+2}=\mathbb{P}_{y_0}^{2n+2}\otimes\mathbb{D}$, where $\mathbb{P}_{y_T}^{2n+1}$ and $\mathbb{P}_{y_0}^{2n+2}$ are path measures on $C([0,T],\mathcal{X})$. 
The time-reversal path measure $(\mathbb{P}_{y_T}^{2n+1})^R$ is induced by (\ref{eqn:IPF_backward}) while $\mathbb{P}_{y_0}^{2n+2}$ is induced by (\ref{eqn:IPF_forward})
\begin{align}
    \rmd Z_t &=f_{T-t}^{2n+1}(Z_t,y_T)\rmd t + \rmd W_t,\quad Z_0\sim \mathcal{N}(0,I), \label{eqn:IPF_backward}\\
    \rmd X_t &=f_t^{2n+2}(X_t,y_0)\rmd t + \rmd B_t,\quad X_0 \sim p(x|y_0),\label{eqn:IPF_forward}
\end{align}
where $(B_t)_{t\in[0,T]}$ and $(W_t)_{t\in[0,T]}$ are $d$-dimensional standard
Brownian motions. Hence $(Z_t)_{t \in [0,T]}$ given by
  \eqref{eqn:IPF_backward} corresponds to a backward process, while
  $(X_t)_{t \in [0,T]}$ given by \eqref{eqn:IPF_forward} corresponds to a forward process. Sampling from
the time-reversal of $\Pi^{2n+1}=\mathbb{P}_{y_T}^{2n+1}\otimes\mathbb{D}$
requires additionally sampling $Y_T \sim p(y)$ to ensure
$(Z_0,Y) \sim \Pi_T(z,y)$ while sampling from
$\Pi^{2n+2}=\mathbb{P}_{y_0}^{2n+2}\otimes\mathbb{D}$ requires sampling
$Y_0 \sim p(y)$ to ensure $(X_0,Y_0) \sim \Pi_0(x,y)$. The above drift functions satisfy the following recursions for $n\in\mathbb{N}_0$:
\begin{align}
    f_t^{2n+1}(x_t,y) &= -f_t^{2n}(x_t,y) + \nabla_{x_t}\log\Pi_t^{2n}(x_t|y),\label{eq:driftrecursionDSBa}\\
    f_t^{2n+2}(x_t,y) &= -f_t^{2n+1}(x_t,y) + \nabla_{x_t}\log\Pi_t^{2n+1}(x_t|y),\label{eq:driftrecursionDSBb}
\end{align}
initialized at $f_t^0(x_t)=-x_t/2$. At iteration $n=0$, under the law
$\mathbb{P}_y^1$, the process \eqref{eqn:IPF_backward} is exactly the denoising
process \eqref{eqn:denoising_SDE} with initialization
$Z_0\sim\mathcal{N}(0,I)$. Hence we can proceed with DSM to approximate
$\nabla_{x_t}\log\Pi_t^{0}(x_t|y)$. This recovers DDPS of Section
\ref{sec:DDPS}; iterating further allows us to improve performance when $T$ is
not sufficiently large.  The next iteration $n=1$ then defines a forward process
\eqref{eqn:IPF_forward} under $\mathbb{P}_y^2$ by time-reversing the backward
process associated to $\mathbb{P}_y^1$ and initializing $(X_0, Y_0)\sim
p(x,y)$. Another application of DSM approximates the next score function
$\nabla_{x_t}\log\Pi_t^{1}(x_t|y)$ by simulating trajectories under
$\mathbb{P}_y^1$ using the previous score approximation. One can avoid storing
all score approximations by adopting the \emph{mean-matching} approach
in \citep{de2021diffusion,shi2022conditional}.  Iterating in this manner until
convergence yields a backward process \eqref{eqn:IPF_backward} that we refer to
as the Diffusion Schr\"{o}dinger Bridge Posterior Sampler (DSB-PS); see Algorithm \ref{alg:DSBPS} for an algorithmic outline, and
\citet[Section 4]{shi2022conditional} for more implementation
details\footnote{Code for DDPS and DSB-PS is available
  \href{https://github.com/vdeborto/cdsb}{here}.}. In practice, the drift
functions $f_t^{2n+1}$ and $f_t^{2n+2}$ are approximated by
$f_{t}^{\theta,2n+1}$ and $f_{t}^{\phi,2n+2}$.

\begin{algorithm}
  \protect\caption{Diffusion Schr\"odinger Bridge for posterior sampling \label{alg:DSBPS}}
  \begin{algorithmic}
\small
\STATE{\textbf{--- Training procedure:}}
\FOR{$n = 0, \dots, N$}
\STATE{\textbf{-- Approximation of $f_t^{2n+1}$ by $f_t^{\theta, 2n+1}$}}
\WHILE{not converged}
\STATE{Sample $(X_t^k)_{t \in [0,T]}$ using  \eqref{eqn:IPF_forward} where $(X_0^k, Y_0^k) \sim p(x,y)$ with drift $f_t^{\phi,2n}$  for $k\in[K]$.} 
\STATE{Set $Y_t^k = Y_0^k$ for $t \in (0,T]$ and $k\in[K]$.}
\STATE{Approximate the mean-matching loss \citep[Section 3.1]{shi2022conditional} using $(X_t^k, Y_t^k)_{t \in [0,T],k\in[K]}$.} 
\STATE{Update $\theta$ in $f_t^{\theta,2n+1}$ using Adam optimizer.}
\ENDWHILE
\STATE{\textbf{-- Approximation of $f_t^{2n+2}$ by $f_t^{\phi, 2n+2}$}}
\WHILE{not converged}
\STATE{Sample $(Z_t^k)_{t \in [0,T]}$ using  \eqref{eqn:IPF_backward} where $(Z_0^k, Y_T^k) \sim \mathcal{N}(x;0,I)p(y)$ with drift $f_t^{\theta, 2n+1}$ for $k\in[K]$.} 
\STATE{Set $Y_t^k = Y_T^k$ for $t \in [0,T)$ and $k\in[K]$.}
\STATE{Approximate the mean-matching loss \citep[Section 3.1]{shi2022conditional} using $(Z_t^k, Y_t^k)_{t \in [0,T],k\in[K]}$.}
\STATE{Update $\phi$ in $f_t^{\phi,2n+2}$ using Adam optimizer.}
\ENDWHILE
\ENDFOR
\STATE{\textbf{--- Sampling procedure:}}
\STATE{Sample $(Z_t)_{t\in[0,T]}$ using the SDE \eqref{eqn:IPF_backward} for the observation $y$.}
\STATE{\textbf{return:} $Z_T$ approximately distributed according to $p(x|y)$.}
  \end{algorithmic}
\end{algorithm}

\section{Denoising diffusions for general sampling\label{sec:DDGS}}
As DDPS and DSB-PS are amortized variational inference procedures, we cannot
expect good performance for observations $y\in\mathcal{Y}$ that are very
unlikely under the model distribution $p(y)$, although this can be partially
mitigated by designing mechanisms to sample synthetic observations closer to the
available ones \citep{sharrock2022sequential}. These samplers are also not
applicable to general distributions that cannot be written as the posterior
distribution of a statistical model and for which we only have access to an unnormalized density instead of samples from the joint model distribution. We now present an
algorithm proposed by \citet{vargasDDSampler2023} that can deal with these two
limitations. Let $p(x)=\gamma(x)/\mathcal{Z}$ denote a target distribution on
$\mathcal{X}$, where $\gamma:\mathcal{X}\rightarrow\mathbb{R}_+$ can
be evaluated pointwise and the normalizing constant
$\mathcal{Z}=\int_{\mathcal{X}}\gamma(x)\rmd x$ is intractable.

Like before, we consider an OU process $(X_t)_{t\in[0,T]}$ \eqref{eqn:noising_SDE} with initialization $X_0\sim p(x)$, and denote the induced path measure on $C([0,T],\mathcal{X})$ as $\mathbb{P}$. The time-reversal $(Z_t)_{t\in[0,T]} = (X_{T-t})_{t\in[0,T]}$ satisfies (weakly) the SDE
\begin{align}\label{eqn:DD_GS_reversal}
    \rmd Z_t = \tfrac{1}{2}Z_t \rmd t+\nabla_{z_t}\log p_{T-t}(Z_t)\rmd t + \rmd W_t,
\end{align}
with $ Z_0\sim p_T(x_T)$ and $ p_t(x_t)=\int_{\mathcal{X}} p_{t|0}(x_t|x_0)p(x_0)\rmd x_0$ denotes the
marginal distribution of $X_t$. In contrast to Section \ref{sec:DDPS}, we cannot
approximate the score function $\nabla_{x_t}\log p_{t}(x_t)$ by minimizing a
loss similar to \eqref{eqn:DSM_loss} as it is infeasible to obtain or generate samples from $p(x)$. 
Hence we shall seek an alternative representation of
$\mathbb{P}$. To do so, we introduce a reference path measure $\mathbb{M}$
induced by a stationary OU process, i.e. \eqref{eqn:noising_SDE} with
$X_0\sim\mathcal{N}(0,I)$. This process is reversible, and its time-reversal
satisfies 
\begin{align}\label{eqn:ref_process}
    \rmd Z_t = -\tfrac{1}{2}Z_t \rmd t + \rmd W_t,\quad Z_0\sim \mathcal{N}(0,I).
\end{align}
For $s<t$, we denote the transition density of \eqref{eqn:ref_process} as
$m_{t|s}(z_t|z_s)$.  Writing
$\mathbb{P}=p\otimes\mathbb{M}_{|0}=\Phi\cdot\mathbb{M}$, where
$\mathbb{M}_{|0}$ denotes the law of $\mathbb{M}$ conditioned on $X_0$ and
$\Phi(x_0)=p(x_0)/\mathcal{N}(x_0;0,I)$, we express $\mathbb{P}^R$ as a
\emph{Doob's $h$-transform} \citep[p. 83]{rogers2000diffusions} of
\eqref{eqn:ref_process} under $\mathbb{M}$
\begin{align}\label{eqn:htransform}
    \rmd Z_t = -\tfrac{1}{2}Z_t \rmd t+\nabla_{z_t}\log h_{T-t}(Z_t)\rmd t + \rmd W_t,
\end{align}
with $ Z_0\sim p_T(x_T)$ and where we have
\begin{equation*}
  \textstyle 
 h_t(x_t)=\int_{\mathcal{X}}\Phi(x_0)m_{T|T-t}(x_0|x_t)\rmd x_0 , 
\end{equation*}
which can be characterized by a backward Kolmogorov equation and satisfies
$\nabla_{x_t}\log h_{t}(x_t)=\nabla_{x_t}\log p_{t}(x_t)+x_t$ from
\eqref{eqn:DD_GS_reversal}. An implementation of \eqref{eqn:htransform} will
involve setting $T$ sufficiently large for $p_T\approx\mathcal{N}(0,I)$, and
approximating $\nabla_{x_t}\log h_t(x_t)$ for all
$(t,x_t)\in[0,T]\times\mathcal{X}$ with a parametric function class
$\{u_t^{\theta}(x_t):\theta\in\Theta\}$. We write $\mathbb{Q}^{\theta}$ for the
law on $C([0,T],\mathcal{X})$ induced by the time-reversal of the diffusion defined by $Z_0\sim \mathcal{N}(0,I)$ and
\begin{align}\label{eqn:KL_proposal}
    \rmd Z_t = -\tfrac{1}{2}Z_t \rmd t+u_{T-t}^{\theta}(Z_t)\rmd t + \rmd W_t.
\end{align}
As we cannot sample from $\mathbb{P}$, we cannot easily obtain unbiased low-variance estimates of the gradient of the forward KL loss $\mathrm{KL}(\mathbb{P}|\mathbb{Q}^{\theta})$ w.r.t. $\theta$. So we instead estimate $\theta$ by minimizing the reverse KL loss \citep[Proposition 1]{vargasDDSampler2023}
\begin{align}\label{eqn:KL_loss}\textstyle
  \mathcal{L}(\theta) &=\mathrm{KL}(\mathbb{Q}^{\theta}|\mathbb{P}) \nonumber \\
  &= \textstyle
    \mathbb{E}_{\mathbb{Q}^{\theta}}[\frac{1}{2}\int_0^T\|u_{T-t}^{\theta}(Z_t)\|^2\rmd t - \log \Phi(Z_T) ].
\end{align}
This is a specific instance of a more general class of KL \emph{optimal control} problems with many connections to sampling \citep{kappen2012optimal,kappen2016adaptive,hengcontrolled2020,zhangyongxinchen2021path}. We note that intractability of the normalizing constant $\mathcal{Z}$ appearing in $\Phi$ is not an issue as it does not affect the minimizers of $\mathcal{L}(\theta)$. 

With a minimizer $\theta$ of \eqref{eqn:KL_loss}, we have a denoising diffusion general sampler (DDGS, Algorithm \ref{alg:DDGS}) that gives approximate samples from $p(x)$ by simulating \eqref{eqn:KL_proposal} and returning $Z_T$, and an unbiased estimator of $\mathcal{Z}$ using importance sampling with proposal law $\mathbb{Q}^{\theta}$ and target law $\mathbb{P}$. An approximate sample from $p(x)$ and an alternative estimator of $\mathcal{Z}$ can be obtained using a \emph{probability flow}, i.e. an ordinary differential equation (ODE) that has the same marginal distributions on $\mathcal{X}$ as $\mathbb{Q}^{\theta}$; see \citet[Section 4.3]{song2020score}, \citet[Section 2.4]{vargas2023bayesian}, and Appendix \ref{app:probaflowODE}. In practice, we have to consider time-discretizations of these continuous-time processes and some care when choosing numerical integrators is necessary \citep[Section 3]{vargasDDSampler2023}. Empirical comparisons between DDGS and Sequential Monte Carlo (SMC) methods can be found in \citet{vargasDDSampler2023}. 

\begin{algorithm}
  \protect\caption{Denoising diffusion for general sampling \label{alg:DDGS}}
  \begin{algorithmic}
\small
\STATE{\textbf{--- Training procedure:}}
\WHILE{not converged}
\STATE{Sample $(Z_t^k)_{t\in[0,T]}$ using the SDE \eqref{eqn:KL_proposal} for $k\in[K]$}.
\STATE{Approximate the loss \eqref{eqn:KL_loss} using $(Z_t^k)_{t \in [0,T],k\in[K]}$.} 
\STATE{Update $\theta$ in $u_t^\theta$ using Adam optimizer.}
\ENDWHILE
\STATE{\textbf{--- Sampling procedure:}}
\STATE{Sample $(Z_t)_{t\in[0,T]}$ using the SDE \eqref{eqn:KL_proposal}.}
\STATE{\textbf{return:} $Z_T$ approximately distributed according to $p(x)$.}
  \end{algorithmic}
\end{algorithm}

\section{Diffusion Schr\"{o}dinger bridges for general sampling}
\label{sec:general_SB}
Analogous to Section \ref{sec:DSB-PS}, we could also improve DDGS by adopting the following SB formulation over probability measures on $C([0,T],\mathcal{X})$:
\begin{align}\label{eqn:SB_GS}
\textstyle
  & \textstyle \Pi^{\star}=\arg\min_{\Pi}\{\mathrm{KL}(\Pi|\mathbb{M}): \nonumber \\
  &  \qquad \Pi_0(x_0)=p(x_0),\quad \Pi_T(x_T)=\mathcal{N}(x_T;0,I)\}.
\end{align}
We refer readers to Appendix \ref{app:SBasentropyOT} and \citet[Section 3.1]{de2021diffusion} for connections to entropy-regularized \emph{optimal transport} problems and a Monge--Kantorovich problem in the zero-noise limit \citep{mikami2004monge,leonard2012schrodinger,leonard2014survey}. 
The IPF recursion solving \eqref{eqn:SB_GS} is
\begin{align}
    &\textstyle \Pi^{2n+1}=\arg\min_{\Pi}\{\mathrm{KL}(\Pi|\Pi^{2n}): \Pi_0(x_0)=p(x_0)\}, \nonumber \\
    &\textstyle \Pi^{2n+2}=\arg\min_{\Pi}\{\mathrm{KL}(\Pi|\Pi^{2n+1} 
): \Pi_T(x_T)=\pi_T(x_T)\}, \nonumber 
\end{align}
for $n\in\mathbb{N}_0$, with $\pi_T(x_T) = \mathcal{N}(x_T;0,I)$ and $\Pi^0=\mathbb{M}$. The path measure  $\Pi^{2n+1}$ is induced by the forward process \eqref{eqn:IPF_GS_forward} while $(\Pi^{2n+2})^R$ is induced by the backward process \eqref{eqn:IPF_GS_backward}
\begin{align}
    \rmd X_t &=f_{t}^{2n+1}(X_t)\rmd t + \rmd B_t,\quad X_0\sim p(x),\label{eqn:IPF_GS_forward}\\
    \rmd Z_t &=f_{T-t}^{2n+2}(Z_t)\rmd t + \rmd W_t,\quad Z_0\sim \mathcal{N}(0,I), \label{eqn:IPF_GS_backward}
\end{align}
with drift functions
\begin{align}\label{eqn:IPF_GS_drift1}
    f_t^{2n+1}(x_t) &= -f_{t}^{2n}(x_t) + \nabla_{x_t}\log\Pi_{t}^{2n}(x_t),\\
    f_t^{2n+2}(x_t) &= f_t^{2n}(x_t) + \nabla_{x_t}\log h_{t}^{2n}(x_t),\label{eqn:IPF_GS_drift2}
\end{align}
initialized at $f_t^0(x_t)=-x_t/2$ and $\Pi_t^0(x_t)=\mathcal{N}(x_t;0,I)$. Here
\begin{equation}\label{eqn:hfunc_general}
  \textstyle 
 h_t^{2n}(x_t)=\int_{\mathcal{X}} \Phi^{2n}(x_0)q_{T|T-t}^{2n}(x_0|x_t)\rmd x_0, 
\end{equation}
where $\Phi^{2n}(x_0)=p(x_0)/\Pi_0^{2n}(x_0)$ and $q_{t|s}^{2n}(z_t|z_s)$ for $s<t$ denotes the transition density of $(Z_t)_{t\in[0,T]}$ under $(\Pi^{2n})^R$. 

Since the first two iterates are $\Pi^1=\mathbb{P}$ defined in Section
\ref{sec:DDGS} and $\Pi^2=\mathcal{N}(0,I)\otimes\mathbb{P}_{|T}$, we recover
DDGS of Section \ref{sec:DDGS} by noticing that
$\nabla_{x_t}\log h_t^0(x_t)=\nabla_{x_t}\log
h_t(x_t)$. 
As $p_T \to \mathcal{N}(0, I)$ when $T \to +\infty$, DDGS provides an
approximation of the SB when the time horizon $T$ is sufficiently large. When
the latter is not the case, further iterations can improve performance. For
$n\geq 1$, $\Pi^{2n+1}=p\otimes\Pi_{|0}^{2n}$ is the law of a forward process
\eqref{eqn:IPF_GS_forward} with a drift function \eqref{eqn:IPF_GS_drift1} that
involves $\nabla_{x_t}\log\Pi_t^{2n}(x_t)$. Simulating from our
approximation of $\Pi^{2n}$, DSM allows us to construct an approximation
$s_t^{\phi,2n}(x_t)$ of this score function. Next by rewriting
$\Pi^{2n+2}=\mathcal{N}(0,I)\otimes\Pi_{|T}^{2n+1}=(h_0^{2n}/h_T^{2n})\cdot\Pi^{2n}$,
we see that its reversal is the law of a backward process
\eqref{eqn:IPF_GS_backward} that is given in terms of a Doob's $h$-transform of
the backward process associated to $(\Pi^{2n})^R$; see Appendix \ref{app:doob}.
To approximate $\nabla_{x_t}\log h_t^{2n}(x_t)$ with $u_t^{\theta,2n}(x_t)$, we
consider another reverse KL minimization
\begin{align}\label{eqn:KL_loss_again}
  \textstyle \mathcal{L}^{2n+2}(\theta)&=\mathrm{KL}(\mathbb{Q}^{\theta}|\Pi^{2n+1}) \nonumber \\
  & \textstyle = 
    \mathbb{E}_{\mathbb{Q}^{\theta}}[\frac{1}{2}\int_0^T\|u_{T-t}^{\theta,2n}(Z_t)\|^2\rmd t - \log \Phi^{2n}(Z_T) ],
\end{align}
where $(\mathbb{Q}^{\theta})^R$ is induced by  the diffusion defined by $Z_0\sim \mathcal{N}(0,I)$ and
\begin{align}\label{eqn:KL_proposal_again}
    \rmd Z_t = f_{T-t}^{2n}(Z_t) \rmd t+u_{T-t}^{\theta,2n}(Z_t)\rmd t + \rmd W_t.
\end{align}
In an implementation of \eqref{eqn:KL_loss_again} and \eqref{eqn:KL_proposal_again}, one has to replace $f_t^{2n}(x_t)$ with the drift approximation from the previous KL minimizations, and approximate $\Pi_0^{2n}(x_0)$ in $\Phi^{2n}(x_0)=p(x_0)/\Pi_0^{2n}(x_0)$ by numerically integrating the probability flow ODE using the score approximation $s_t^{\phi,2n}(x_t)$. One should iterate the above steps until convergence and avoid storing all approximations in memory. The final backward process \eqref{eqn:IPF_GS_backward} yields the diffusion Schr\"{o}dinger bridge general sampler (DSB-GS), outlined in Algorithm \ref{alg:DSBGS}.  During the training part of Algorithm \ref{alg:DSBGS}, we can skip the approximation of $\nabla \log \Pi_t^{2n}$ when $n=0$, since in that case $\Pi_t^{0} = \mathcal{N}(0, I)$. Doing so, we see that the first iteration of Algorithm \ref{alg:DSBGS} recovers Algorithm \ref{alg:DDGS}. 

An alternative Schr\"{o}dinger bridge formulation to sampling has been
formulated in
\citep{follmer1984entropy,daipra1991stochastic,tzen2019theoretical}. This corresponds to (\ref{eqn:SB_GS}) with the reference measure $\mathbb{M}$ defined using a pinned Brownian motion running backwards in time, and the terminal distribution $\Pi_T(x_T)$ given by a Dirac measure at the origin. 
The appealing feature of this simpler SB is that IPF converges in two iterations; various numerical approximations have been developed for this case 
\citep{barrlamacraft2020quantum,zhangmarzouk2021sampling,zhangyongxinchen2021path,vargas2023bayesian}. 
However,
it was observed in \citet{vargasDDSampler2023} that the resulting scores one
needs to estimate are very steep for times close to $T$ due to the degenerate terminal distribution. As a result, this approach can be numerically quite unstable.  

 \begin{algorithm}
  \protect\caption{Diffusion Schr\"odinger Bridge for general sampling \label{alg:DSBGS}}
  \begin{algorithmic}
\small
\STATE{\textbf{--- Training procedure:}}
\FOR{$n = 0, \dots, N$}
\STATE{\textbf{-- Approximation of $\nabla \log \Pi_t^{2n}$ by $s_t^{\phi, 2n}$}}
\WHILE{not converged}
\STATE{Sample $(Z_t^k)_{t \in [0,T]}$ using \eqref{eqn:IPF_GS_backward} with  $f_t^{2n}$ for $k\in[K]$.}
\STATE{Approximate the DSM loss using $(Z_t^k)_{t\in[0,T],k\in[K]}$}
\STATE{Update $\phi$ in $s_t^{\phi,2n}$ using Adam optimizer.}
\ENDWHILE
\STATE{\textbf{-- Approximation of $\nabla \log h_t^{2n}$ by $u_t^{\theta, 2n}$}}
\WHILE{not converged}
\STATE{Sample $(Z_t^k)_{t\in[0,T]}$ using the SDE \eqref{eqn:KL_proposal_again} for $k\in[K]$.}
\STATE{Run probability flow ODE with $f_t^{2n}-\tfrac{1}{2}s_t^{\phi,2n}$ to compute $\Pi_0^{2n}(Z_T^k)$ for $k\in[K]$.}
\STATE{Approximate the KL loss \eqref{eqn:KL_loss_again} using $(Z_t^k)_{t \in [0,T],k\in[K]}$.}
\STATE{Update $\theta$ in $u_t^{\theta, 2n}$ using Adam optimizer.}
\ENDWHILE
\STATE{\textbf{-- Update drift $f_t^{2n+2}$}}
\STATE{Set $f_t^{2n+2} = f_t^{2n} + u_{t}^{\theta, 2n}$ as in \eqref{eqn:IPF_GS_drift2}.}
\ENDFOR
\STATE{\textbf{--- Sampling procedure:}}
\STATE{Sample $(Z_t)_{t\in[0,T]}$ using the SDE \eqref{eqn:IPF_GS_backward}.}
\STATE{\textbf{return:} $Z_T$ approximately distributed according to $p(x)$.}
  \end{algorithmic}
\end{algorithm}

\section{Discussion\label{sec:discuss}}
We have presented a concise overview of how diffusion processes can be used to sample approximately from posterior distributions and any general target distributions. These methods have been successfully used to solve a wide class of sampling problems, and are alternatives to standard MCMC, SMC, and ABC techniques. However, compared to these well-established methods, while there are convergence results available for such diffusion-based samplers \citep{de2022convergence,ChenChewiSinhosamplingaseasyaslearningscore2022}, these results are based on assumptions about the score function estimation error which remain difficult to verify. Promising methodological alternatives to diffusion approaches have also been recently proposed where one constructs a process between two distributions one can sample from, but this process is built using an ordinary differential equation whose drift function can be approximated by solving a simple regression problem \citep{albergo2022building,lipman2022flow,liu2022flow}. The techniques developed in these works have been further extended to compute the SB \citep{shi2023diffusion,peluchetti2023diffusion} and provide an alternative to DSB-PS. 


\begin{appendix}

\section{Probability Flow ODE}\label{app:probaflowODE}
Consider the following SDE on $\mathcal{X}$
\begin{align}\label{eqn:generalSDE}
    \rmd X_t = f_t(X_t)\rmd t + \rmd B_t, \quad X_0 \sim p_0(x),
\end{align}
where $(B_t)_{t\in[0,T]}$ is a standard $d$-dimensional Brownian motion. The marginal density $p_t$ of $X_t$ satisfies the Fokker--Planck--Kolmogorov equation
\begin{align}\label{eq:FPK1}
\partial_t p_t(x) &=-\mathrm{div}(f_t p_t)(x) +\tfrac{1}{2}\Delta p_t(x)\\
&=- \mathrm{div} (\bar{f}_t p_t)(x) \label{eq:FPKflow}
\end{align}
where $\mathrm{div}(u)(x):=\sum_{i=1}^d \tfrac{\partial u_i(x)}{\partial x_i}$ for differentiable $u:\mathcal{X} \rightarrow \mathcal{X}$, $\Delta a(x):=\sum_{i=1}^d \tfrac{\partial^2 a(x)}{\partial x_i^2}$ for twice differentiable $a:\mathcal{X} \rightarrow \mathbb{R}$, and
\begin{equation}
\bar{f}_t(x)=f_t(x)-\tfrac{1}{2}\nabla \log p_t(x).
\end{equation}
Equation \eqref{eq:FPKflow} shows that the ODE 
\begin{align}\label{eqn:probaflowODE}
   \rmd X_t = \bar{f}_t(X_t) \rmd t,\quad X_0 \sim p_0(x),
\end{align}
known as the \emph{probability flow} ODE, is such that $X_t \sim p_t(x)$ for all $t\in[0,T]$, i.e. it admits the same marginal distributions as the SDE in \eqref{eqn:generalSDE}.
By using the instantaneous change of variables formula \citep{chen2018neural}, we obtain
\begin{equation}\label{eq:changeofvarformula}
  \textstyle
\log p_0(x_0)=\log p_T(x_T)+ \int_0^T \mathrm{div}(\bar{f}_t)(x_t) \rmd t,
\end{equation}
for $(x_t)_{t\in[0,T]}$ obtained by solving (\ref{eqn:probaflowODE}) initialized at $X_0=x_0$.

\section{Schr\"odinger Bridge as Entropy-Regularized Optimal Transport}\label{app:SBasentropyOT}
Consider the following generic SB problem over probability measures on $C([0,T],\mathcal{X})$:
\begin{align}\label{eqn:SB_gen}
\textstyle
  & \textstyle \Pi^{\star}=\arg\min_{\Pi}\{\mathrm{KL}(\Pi|\mathbb{S}): \nonumber \\
  &  \qquad \Pi_0(x_0)=\nu_0(x_0),\quad \Pi_T(x_T)=\nu_T(x_T)\},
\end{align}
where $\mathbb{S}$ is a generic path measure on $C([0,T],\mathcal{X})$. We note that the dynamic formulation \eqref{eqn:SB_gen} admits a static analogue. Let $s_{0,T}$ denote the marginal distribution of $(X_0,X_T)$ under $\mathbb{S}$, and $\mathbb{S}_{|0,T}$ denote the law $\mathbb{S}$ conditioned on $(X_0,X_T)$, i.e. we have $\mathbb{S}=s_{0,T}\otimes\mathbb{S}_{|0,T}$. By decomposing the KL divergence in \eqref{eqn:SB_gen}, we see that $\Pi^{\star}=\pi_{0,T}^{\star}\otimes\mathbb{S}_{|0,T}$, where the static SB $\pi_{0,T}^{\star}$ is defined by minimizing over probability measures on $\mathcal{X}\times\mathcal{X}$ with the same marginal constraints
\begin{align}
\textstyle
    \pi_{0,T}^{\star}&=\textstyle \arg\min_{\pi_{0,T}}\{\mathrm{KL}(\pi_{0,T}|s_{0,T}): \nonumber \\
   &\pi_0(x_0)=\nu_0(x_0),\quad \pi_T(x_T)=\nu_T(x_T)\}.
\end{align}
We can rewrite  
\begin{equation*}
\mathrm{KL}(\pi_{0,T}|s_{0,T})=-\mathbb{E}_{\pi_{0,T}}[\log s_{T|0}(X_T|X_0)]-\text{H}(\pi_{0,T}),
\end{equation*}
where $s_{T|0}(x_T|x_0)$ denotes the transition density under $\mathbb{S}$ and $\text{H}(\pi_{0,T})=-\mathbb{E}_{\pi_{0,T}}[\log\pi_{0,T}(X_0,X_T)]$ is an entropy term. In particular, if $\mathbb{S}$ is the law of a standard Brownian motion on $\mathcal{X}$, we have 
\begin{align}
\textstyle
    \pi_{0,T}^{\star}&=\textstyle \arg\min_{\pi_{0,T}}\{\mathbb{E}_{\pi_{0,T}}[||X_T-X_0||^2]-2T\text{H}(\pi_{0,T}): \nonumber \\
   &\pi_0(x_0)=\nu_0(x_0),\quad \pi_T(x_T)=\nu_T(x_T)\}\nonumber.
\end{align}
The above can be seen as an entropy-regularized \emph{optimal transport} problem that converges to a Monge--Kantorovich problem as $T \rightarrow 0$ \citep{mikami2004monge,leonard2012schrodinger,leonard2014survey}. 

\section{Iterative proportional fitting as Doob's $h$-transforms\label{app:doob}}
Recall from Section \ref{sec:general_SB} that the IPF iterates satisfy
\begin{align}
    \Pi^{2n+1}=p\otimes\Pi_{|0}^{2n},\quad 
    \Pi^{2n+2}=\mathcal{N}(0,I)\otimes\Pi_{|T}^{2n+1}
\end{align}
for $n\in\mathbb{N}_0$, which implies 
\begin{align}\label{eqn:even_IPF_updates}
    \Pi^{2n+2}=\frac{p(x_0)\mathcal{N}(x_T;0,I)}{\Pi_0^{2n}(x_0)\Pi_T^{2n+1}(x_T)}\Pi^{2n}.
\end{align}
From the definition of $h_t^{2n}(x_t)$ in \eqref{eqn:hfunc_general}, we note that 
\begin{align}\label{eqn:IPF_gen_0}
    h_0^{2n}(x_0)=\Phi^{2n}(x_0)=p(x_0)/\Pi_0^{2n}(x_0),
\end{align}
and 
\begin{align}
\textstyle
    h_T^{2n}(x_T)=\int_{\mathcal{X}}\Phi^{2n}(x_0)q_{T|0}^{2n}(x_0|x_T)\rmd x_0.
\end{align}
Denoting the conditional distribution of $X_T$ given $X_0$  under $\Pi^{2n}$ as $\pi^{2n}(x_T|x_0)$, we have 
\begin{align}
\textstyle
    \Pi_T^{2n+1}(x_T) &= \textstyle\int_{\mathcal{X}}p(x_0)\pi^{2n}(x_T|x_0)\rmd x_0\nonumber\\
    &\textstyle= \mathcal{N}(x_T;0,I)\int_{\mathcal{X}}\Phi^{2n}(x_0)q_{T|0}^{2n}(x_0|x_T)\rmd x_0,
\end{align}
which gives 
\begin{align}\label{eqn:IPF_gen_1}
    \frac{\mathcal{N}(x_T;0,I)}{\Pi_T^{2n+1}(x_T)} = h_T^{2n}(x_T)^{-1}.
\end{align}
Combining \eqref{eqn:even_IPF_updates}, \eqref{eqn:IPF_gen_0}, and \eqref{eqn:IPF_gen_1} gives 
\begin{align}
    \Pi^{2n+2} = \frac{h_0^{2n}(x_0)}{h_T^{2n}(x_T)}\Pi^{2n}.
\end{align}

\end{appendix}


\begin{funding}
A.D. is partially supported by EPSRC grants EP/R034710/1 (CoSinES) and EP/R018561/1 (Bayes4Health). J.H. was funded by CY Initiative of Excellence (grant ``Investissements d'Avenir'' ANR-16-IDEX-0008). 
\end{funding}

\bibliographystyle{imsart-nameyear} 
\bibliography{ref}       

\begin{thebibliography}{62}

\bibitem[\protect\citeauthoryear{Albergo and
  Vanden{-}Eijnden}{2023}]{albergo2022building}
\begin{binproceedings}[author]
\bauthor{\bsnm{Albergo},~\bfnm{Michael~S.}\binits{M.~S.}} \AND
  \bauthor{\bsnm{Vanden{-}Eijnden},~\bfnm{Eric}\binits{E.}}
(\byear{2023}).
\btitle{Building Normalizing Flows with Stochastic Interpolants}.
In \bbooktitle{The Eleventh International Conference on Learning
  Representations, {ICLR} 2023, Kigali, Rwanda, May 1-5, 2023}.
\bpublisher{OpenReview.net}.
\end{binproceedings}
\endbibitem

\bibitem[\protect\citeauthoryear{Anderson}{1982}]{anderson1982reverse}
\begin{barticle}[author]
\bauthor{\bsnm{Anderson},~\bfnm{Brian~DO}\binits{B.~D.}}
(\byear{1982}).
\btitle{Reverse-time diffusion equation models}.
\bjournal{Stochastic Processes and Their Applications}
\bvolume{12}
\bpages{313--326}.
\end{barticle}
\endbibitem

\bibitem[\protect\citeauthoryear{Austin et~al.}{2021}]{austin2021structured}
\begin{binproceedings}[author]
\bauthor{\bsnm{Austin},~\bfnm{Jacob}\binits{J.}},
  \bauthor{\bsnm{Johnson},~\bfnm{Daniel~D.}\binits{D.~D.}},
  \bauthor{\bsnm{Ho},~\bfnm{Jonathan}\binits{J.}},
  \bauthor{\bsnm{Tarlow},~\bfnm{Daniel}\binits{D.}} \AND
  \bauthor{\bparticle{van~den} \bsnm{Berg},~\bfnm{Rianne}\binits{R.}}
(\byear{2021}).
\btitle{Structured Denoising Diffusion Models in Discrete State-Spaces}.
In \bbooktitle{Advances in Neural Information Processing Systems 34: Annual
  Conference on Neural Information Processing Systems 2021, NeurIPS 2021,
  December 6-14, 2021, virtual}
(\beditor{\bfnm{Marc'Aurelio}\binits{M.}~\bsnm{Ranzato}},
  \beditor{\bfnm{Alina}\binits{A.}~\bsnm{Beygelzimer}},
  \beditor{\bfnm{Yann~N.}\binits{Y.~N.}~\bsnm{Dauphin}},
  \beditor{\bfnm{Percy}\binits{P.}~\bsnm{Liang}} \AND
  \beditor{\bfnm{Jennifer~Wortman}\binits{J.~W.}~\bsnm{Vaughan}}, eds.)
\bpages{17981--17993}.
\end{binproceedings}
\endbibitem

\bibitem[\protect\citeauthoryear{Barr, Gispen and
  Lamacraft}{2020}]{barrlamacraft2020quantum}
\begin{binproceedings}[author]
\bauthor{\bsnm{Barr},~\bfnm{Ariel}\binits{A.}},
  \bauthor{\bsnm{Gispen},~\bfnm{Willem}\binits{W.}} \AND
  \bauthor{\bsnm{Lamacraft},~\bfnm{Austen}\binits{A.}}
(\byear{2020}).
\btitle{Quantum ground states from reinforcement learning}.
In \bbooktitle{Mathematical and Scientific Machine Learning}.
\end{binproceedings}
\endbibitem

\bibitem[\protect\citeauthoryear{Beaumont}{2019}]{beaumont2019approximate}
\begin{barticle}[author]
\bauthor{\bsnm{Beaumont},~\bfnm{Mark~A}\binits{M.~A.}}
(\byear{2019}).
\btitle{Approximate {B}ayesian computation}.
\bjournal{Annual Review of Statistics and Its Application}
\bvolume{6}
\bpages{379--403}.
\end{barticle}
\endbibitem

\bibitem[\protect\citeauthoryear{Benton et~al.}{2022}]{benton2022denoising}
\begin{barticle}[author]
\bauthor{\bsnm{Benton},~\bfnm{Joe}\binits{J.}},
  \bauthor{\bsnm{Shi},~\bfnm{Yuyang}\binits{Y.}},
  \bauthor{\bsnm{De~Bortoli},~\bfnm{Valentin}\binits{V.}},
  \bauthor{\bsnm{Deligiannidis},~\bfnm{George}\binits{G.}} \AND
  \bauthor{\bsnm{Doucet},~\bfnm{Arnaud}\binits{A.}}
(\byear{2022}).
\btitle{From denoising diffusions to denoising {M}arkov models}.
\bjournal{arXiv preprint arXiv:2211.03595}.
\end{barticle}
\endbibitem

\bibitem[\protect\citeauthoryear{Campbell
  et~al.}{2022}]{campbell2022continuous}
\begin{binproceedings}[author]
\bauthor{\bsnm{Campbell},~\bfnm{Andrew}\binits{A.}},
  \bauthor{\bsnm{Benton},~\bfnm{Joe}\binits{J.}},
  \bauthor{\bsnm{De~Bortoli},~\bfnm{Valentin}\binits{V.}},
  \bauthor{\bsnm{Rainforth},~\bfnm{Tom}\binits{T.}},
  \bauthor{\bsnm{Deligiannidis},~\bfnm{George}\binits{G.}} \AND
  \bauthor{\bsnm{Doucet},~\bfnm{Arnaud}\binits{A.}}
(\byear{2022}).
\btitle{A Continuous Time Framework for Discrete Denoising Models}.
In \bbooktitle{Advances in Neural Information Processing Systems 35: Annual
  Conference on Neural Information Processing Systems 2022, NeurIPS 2022}
(\beditor{\bfnm{S.}\binits{S.}~\bsnm{Koyejo}},
  \beditor{\bfnm{S.}\binits{S.}~\bsnm{Mohamed}},
  \beditor{\bfnm{A.}\binits{A.}~\bsnm{Agarwal}},
  \beditor{\bfnm{D.}\binits{D.}~\bsnm{Belgrave}},
  \beditor{\bfnm{K.}\binits{K.}~\bsnm{Cho}} \AND
  \beditor{\bfnm{A.}\binits{A.}~\bsnm{Oh}}, eds.).
\end{binproceedings}
\endbibitem

\bibitem[\protect\citeauthoryear{Chen, Georgiou and
  Pavon}{2016}]{chen2016entropic}
\begin{barticle}[author]
\bauthor{\bsnm{Chen},~\bfnm{Yongxin}\binits{Y.}},
  \bauthor{\bsnm{Georgiou},~\bfnm{Tryphon}\binits{T.}} \AND
  \bauthor{\bsnm{Pavon},~\bfnm{Michele}\binits{M.}}
(\byear{2016}).
\btitle{Entropic and displacement interpolation: a computational approach using
  the {H}ilbert metric}.
\bjournal{SIAM Journal on Applied Mathematics}
\bvolume{76}
\bpages{2375--2396}.
\end{barticle}
\endbibitem

\bibitem[\protect\citeauthoryear{Chen, Liu and
  Theodorou}{2022}]{chenlikelihood}
\begin{binproceedings}[author]
\bauthor{\bsnm{Chen},~\bfnm{Tianrong}\binits{T.}},
  \bauthor{\bsnm{Liu},~\bfnm{Guan{-}Horng}\binits{G.}} \AND
  \bauthor{\bsnm{Theodorou},~\bfnm{Evangelos~A.}\binits{E.~A.}}
(\byear{2022}).
\btitle{Likelihood Training of Schr{\"{o}}dinger Bridge using Forward-Backward
  SDEs Theory}.
In \bbooktitle{The Tenth International Conference on Learning Representations,
  {ICLR} 2022, Virtual Event, April 25-29, 2022}.
\bpublisher{OpenReview.net}.
\end{binproceedings}
\endbibitem

\bibitem[\protect\citeauthoryear{Chen et~al.}{2018}]{chen2018neural}
\begin{binproceedings}[author]
\bauthor{\bsnm{Chen},~\bfnm{Tian~Qi}\binits{T.~Q.}},
  \bauthor{\bsnm{Rubanova},~\bfnm{Yulia}\binits{Y.}},
  \bauthor{\bsnm{Bettencourt},~\bfnm{Jesse}\binits{J.}} \AND
  \bauthor{\bsnm{Duvenaud},~\bfnm{David}\binits{D.}}
(\byear{2018}).
\btitle{Neural Ordinary Differential Equations}.
In \bbooktitle{Advances in Neural Information Processing Systems 31: Annual
  Conference on Neural Information Processing Systems 2018, NeurIPS 2018,
  December 3-8, 2018, Montr{\'{e}}al, Canada}
(\beditor{\bfnm{Samy}\binits{S.}~\bsnm{Bengio}},
  \beditor{\bfnm{Hanna~M.}\binits{H.~M.}~\bsnm{Wallach}},
  \beditor{\bfnm{Hugo}\binits{H.}~\bsnm{Larochelle}},
  \beditor{\bfnm{Kristen}\binits{K.}~\bsnm{Grauman}},
  \beditor{\bfnm{Nicol{\`{o}}}\binits{N.}~\bsnm{Cesa{-}Bianchi}} \AND
  \beditor{\bfnm{Roman}\binits{R.}~\bsnm{Garnett}}, eds.)
\bpages{6572--6583}.
\end{binproceedings}
\endbibitem

\bibitem[\protect\citeauthoryear{Chen
  et~al.}{2023}]{ChenChewiSinhosamplingaseasyaslearningscore2022}
\begin{binproceedings}[author]
\bauthor{\bsnm{Chen},~\bfnm{Sitan}\binits{S.}},
  \bauthor{\bsnm{Chewi},~\bfnm{Sinho}\binits{S.}},
  \bauthor{\bsnm{Li},~\bfnm{Jerry}\binits{J.}},
  \bauthor{\bsnm{Li},~\bfnm{Yuanzhi}\binits{Y.}},
  \bauthor{\bsnm{Salim},~\bfnm{Adil}\binits{A.}} \AND
  \bauthor{\bsnm{Zhang},~\bfnm{Anru}\binits{A.}}
(\byear{2023}).
\btitle{Sampling is as easy as learning the score: theory for diffusion models
  with minimal data assumptions}.
In \bbooktitle{The Eleventh International Conference on Learning
  Representations, {ICLR} 2023, Kigali, Rwanda, May 1-5, 2023}.
\bpublisher{OpenReview.net}.
\end{binproceedings}
\endbibitem

\bibitem[\protect\citeauthoryear{Chung et~al.}{2023}]{chung2209diffusion}
\begin{binproceedings}[author]
\bauthor{\bsnm{Chung},~\bfnm{Hyungjin}\binits{H.}},
  \bauthor{\bsnm{Kim},~\bfnm{Jeongsol}\binits{J.}},
  \bauthor{\bsnm{McCann},~\bfnm{Michael~Thompson}\binits{M.~T.}},
  \bauthor{\bsnm{Klasky},~\bfnm{Marc~Louis}\binits{M.~L.}} \AND
  \bauthor{\bsnm{Ye},~\bfnm{Jong~Chul}\binits{J.~C.}}
(\byear{2023}).
\btitle{Diffusion Posterior Sampling for General Noisy Inverse Problems}.
In \bbooktitle{The Eleventh International Conference on Learning
  Representations, {ICLR} 2023, Kigali, Rwanda, May 1-5, 2023}.
\bpublisher{OpenReview.net}.
\end{binproceedings}
\endbibitem

\bibitem[\protect\citeauthoryear{Dai~Pra}{1991}]{daipra1991stochastic}
\begin{barticle}[author]
\bauthor{\bsnm{Dai~Pra},~\bfnm{Paolo}\binits{P.}}
(\byear{1991}).
\btitle{A stochastic control approach to reciprocal diffusion processes}.
\bjournal{Applied Mathematics and Optimization}
\bvolume{23}
\bpages{313--329}.
\end{barticle}
\endbibitem

\bibitem[\protect\citeauthoryear{De~Bortoli}{2022}]{de2022convergence}
\begin{barticle}[author]
\bauthor{\bsnm{De~Bortoli},~\bfnm{Valentin}\binits{V.}}
(\byear{2022}).
\btitle{Convergence of denoising diffusion models under the manifold
  hypothesis}.
\bjournal{Transactions on Machine Learning Research}
\bvolume{2022}.
\end{barticle}
\endbibitem

\bibitem[\protect\citeauthoryear{De~Bortoli et~al.}{2021}]{de2021diffusion}
\begin{binproceedings}[author]
\bauthor{\bsnm{De~Bortoli},~\bfnm{Valentin}\binits{V.}},
  \bauthor{\bsnm{Thornton},~\bfnm{James}\binits{J.}},
  \bauthor{\bsnm{Heng},~\bfnm{Jeremy}\binits{J.}} \AND
  \bauthor{\bsnm{Doucet},~\bfnm{Arnaud}\binits{A.}}
(\byear{2021}).
\btitle{Diffusion Schr{\"{o}}dinger Bridge with Applications to Score-Based
  Generative Modeling}.
In \bbooktitle{Advances in Neural Information Processing Systems 34: Annual
  Conference on Neural Information Processing Systems 2021, NeurIPS 2021,
  December 6-14, 2021, virtual}
(\beditor{\bfnm{Marc'Aurelio}\binits{M.}~\bsnm{Ranzato}},
  \beditor{\bfnm{Alina}\binits{A.}~\bsnm{Beygelzimer}},
  \beditor{\bfnm{Yann~N.}\binits{Y.~N.}~\bsnm{Dauphin}},
  \beditor{\bfnm{Percy}\binits{P.}~\bsnm{Liang}} \AND
  \beditor{\bfnm{Jennifer~Wortman}\binits{J.~W.}~\bsnm{Vaughan}}, eds.)
\bpages{17695--17709}.
\end{binproceedings}
\endbibitem

\bibitem[\protect\citeauthoryear{De~Bortoli et~al.}{2022}]{de2022riemannian}
\begin{binproceedings}[author]
\bauthor{\bsnm{De~Bortoli},~\bfnm{Valentin}\binits{V.}},
  \bauthor{\bsnm{Mathieu},~\bfnm{Emile}\binits{E.}},
  \bauthor{\bsnm{Hutchinson},~\bfnm{Michael}\binits{M.}},
  \bauthor{\bsnm{Thornton},~\bfnm{James}\binits{J.}},
  \bauthor{\bsnm{Teh},~\bfnm{Yee~Whye}\binits{Y.~W.}} \AND
  \bauthor{\bsnm{Doucet},~\bfnm{Arnaud}\binits{A.}}
(\byear{2022}).
\btitle{Riemannian score-based generative modeling}.
In \bbooktitle{Advances in Neural Information Processing Systems 35: Annual
  Conference on Neural Information Processing Systems 2022, NeurIPS 2022}
(\beditor{\bfnm{S.}\binits{S.}~\bsnm{Koyejo}},
  \beditor{\bfnm{S.}\binits{S.}~\bsnm{Mohamed}},
  \beditor{\bfnm{A.}\binits{A.}~\bsnm{Agarwal}},
  \beditor{\bfnm{D.}\binits{D.}~\bsnm{Belgrave}},
  \beditor{\bfnm{K.}\binits{K.}~\bsnm{Cho}} \AND
  \beditor{\bfnm{A.}\binits{A.}~\bsnm{Oh}}, eds.).
\end{binproceedings}
\endbibitem

\bibitem[\protect\citeauthoryear{Deming and Stephan}{1940}]{deming1940least}
\begin{barticle}[author]
\bauthor{\bsnm{Deming},~\bfnm{W~Edwards}\binits{W.~E.}} \AND
  \bauthor{\bsnm{Stephan},~\bfnm{Frederick~F}\binits{F.~F.}}
(\byear{1940}).
\btitle{On a least squares adjustment of a sampled frequency table when the
  expected marginal totals are known}.
\bjournal{The Annals of Mathematical Statistics}
\bvolume{11}
\bpages{427--444}.
\end{barticle}
\endbibitem

\bibitem[\protect\citeauthoryear{Dockhorn, Vahdat and
  Kreis}{2022}]{dockhorn2021score}
\begin{binproceedings}[author]
\bauthor{\bsnm{Dockhorn},~\bfnm{Tim}\binits{T.}},
  \bauthor{\bsnm{Vahdat},~\bfnm{Arash}\binits{A.}} \AND
  \bauthor{\bsnm{Kreis},~\bfnm{Karsten}\binits{K.}}
(\byear{2022}).
\btitle{Score-Based Generative Modeling with Critically-Damped Langevin
  Diffusion}.
In \bbooktitle{The Tenth International Conference on Learning Representations,
  {ICLR} 2022, Virtual Event, April 25-29, 2022}.
\bpublisher{OpenReview.net}.
\end{binproceedings}
\endbibitem

\bibitem[\protect\citeauthoryear{F\"ollmer}{1984}]{follmer1984entropy}
\begin{barticle}[author]
\bauthor{\bsnm{F\"ollmer},~\bfnm{Hans}\binits{H.}}
(\byear{1984}).
\btitle{An entropy approach to the time reversal of diffusion processes}.
\bjournal{Lecture Notes in Control and Information Sciences}
\bvolume{69}
\bpages{156--163}.
\end{barticle}
\endbibitem

\bibitem[\protect\citeauthoryear{Fortet}{1940}]{fortet1940resolution}
\begin{barticle}[author]
\bauthor{\bsnm{Fortet},~\bfnm{Robert}\binits{R.}}
(\byear{1940}).
\btitle{R{\'e}solution d’un syst{\`e}me d’{\'e}quations de {M}.
  {S}chr{\"o}dinger}.
\bjournal{Journal de Math{\'e}matiques Pures et Appliqu{\'e}es}
\bvolume{1}
\bpages{83--105}.
\end{barticle}
\endbibitem

\bibitem[\protect\citeauthoryear{Grenander and
  Miller}{1994}]{grenander1994representations}
\begin{barticle}[author]
\bauthor{\bsnm{Grenander},~\bfnm{Ulf}\binits{U.}} \AND
  \bauthor{\bsnm{Miller},~\bfnm{Michael~I}\binits{M.~I.}}
(\byear{1994}).
\btitle{Representations of knowledge in complex systems}.
\bjournal{Journal of the Royal Statistical Society: Series B (Methodological)}
\bvolume{56}
\bpages{549--581}.
\end{barticle}
\endbibitem

\bibitem[\protect\citeauthoryear{Haussmann and
  Pardoux}{1986}]{haussmann1986time}
\begin{barticle}[author]
\bauthor{\bsnm{Haussmann},~\bfnm{Ulrich~G}\binits{U.~G.}} \AND
  \bauthor{\bsnm{Pardoux},~\bfnm{Etienne}\binits{E.}}
(\byear{1986}).
\btitle{Time reversal of diffusions}.
\bjournal{The Annals of Probability}
\bpages{1188--1205}.
\end{barticle}
\endbibitem

\bibitem[\protect\citeauthoryear{Heng et~al.}{2020}]{hengcontrolled2020}
\begin{barticle}[author]
\bauthor{\bsnm{Heng},~\bfnm{Jeremy}\binits{J.}},
  \bauthor{\bsnm{Bishop},~\bfnm{Adrian~N.}\binits{A.~N.}},
  \bauthor{\bsnm{Deligiannidis},~\bfnm{George}\binits{G.}} \AND
  \bauthor{\bsnm{Doucet},~\bfnm{Arnaud}\binits{A.}}
(\byear{2020}).
\btitle{{Controlled sequential Monte Carlo}}.
\bjournal{The Annals of Statistics}
\bvolume{48}
\bpages{2904 -- 2929}.
\end{barticle}
\endbibitem

\bibitem[\protect\citeauthoryear{Ho, Jain and Abbeel}{2020}]{ho2020denoising}
\begin{binproceedings}[author]
\bauthor{\bsnm{Ho},~\bfnm{Jonathan}\binits{J.}},
  \bauthor{\bsnm{Jain},~\bfnm{Ajay}\binits{A.}} \AND
  \bauthor{\bsnm{Abbeel},~\bfnm{Pieter}\binits{P.}}
(\byear{2020}).
\btitle{Denoising Diffusion Probabilistic Models}.
In \bbooktitle{Advances in Neural Information Processing Systems 33: Annual
  Conference on Neural Information Processing Systems 2020, NeurIPS 2020,
  December 6-12, 2020, virtual}
(\beditor{\bfnm{Hugo}\binits{H.}~\bsnm{Larochelle}},
  \beditor{\bfnm{Marc'Aurelio}\binits{M.}~\bsnm{Ranzato}},
  \beditor{\bfnm{Raia}\binits{R.}~\bsnm{Hadsell}},
  \beditor{\bfnm{Maria{-}Florina}\binits{M.}~\bsnm{Balcan}} \AND
  \beditor{\bfnm{Hsuan{-}Tien}\binits{H.}~\bsnm{Lin}}, eds.).
\end{binproceedings}
\endbibitem

\bibitem[\protect\citeauthoryear{Hoogeboom et~al.}{2021}]{hoogeboom2021argmax}
\begin{binproceedings}[author]
\bauthor{\bsnm{Hoogeboom},~\bfnm{Emiel}\binits{E.}},
  \bauthor{\bsnm{Nielsen},~\bfnm{Didrik}\binits{D.}},
  \bauthor{\bsnm{Jaini},~\bfnm{Priyank}\binits{P.}},
  \bauthor{\bsnm{Forr{\'{e}}},~\bfnm{Patrick}\binits{P.}} \AND
  \bauthor{\bsnm{Welling},~\bfnm{Max}\binits{M.}}
(\byear{2021}).
\btitle{Argmax Flows and Multinomial Diffusion: Learning Categorical
  Distributions}.
In \bbooktitle{Advances in Neural Information Processing Systems 34: Annual
  Conference on Neural Information Processing Systems 2021, NeurIPS 2021,
  December 6-14, 2021, virtual}
(\beditor{\bfnm{Marc'Aurelio}\binits{M.}~\bsnm{Ranzato}},
  \beditor{\bfnm{Alina}\binits{A.}~\bsnm{Beygelzimer}},
  \beditor{\bfnm{Yann~N.}\binits{Y.~N.}~\bsnm{Dauphin}},
  \beditor{\bfnm{Percy}\binits{P.}~\bsnm{Liang}} \AND
  \beditor{\bfnm{Jennifer~Wortman}\binits{J.~W.}~\bsnm{Vaughan}}, eds.)
\bpages{12454--12465}.
\end{binproceedings}
\endbibitem

\bibitem[\protect\citeauthoryear{Huang et~al.}{2022}]{huang2022riemannian}
\begin{binproceedings}[author]
\bauthor{\bsnm{Huang},~\bfnm{Chin{-}Wei}\binits{C.}},
  \bauthor{\bsnm{Aghajohari},~\bfnm{Milad}\binits{M.}},
  \bauthor{\bsnm{Bose},~\bfnm{Joey}\binits{J.}},
  \bauthor{\bsnm{Panangaden},~\bfnm{Prakash}\binits{P.}} \AND
  \bauthor{\bsnm{Courville},~\bfnm{Aaron~C.}\binits{A.~C.}}
(\byear{2022}).
\btitle{Riemannian Diffusion Models}.
In \bbooktitle{Advances in Neural Information Processing Systems 35: Annual
  Conference on Neural Information Processing Systems 2022, NeurIPS 2022}
(\beditor{\bfnm{S.}\binits{S.}~\bsnm{Koyejo}},
  \beditor{\bfnm{S.}\binits{S.}~\bsnm{Mohamed}},
  \beditor{\bfnm{A.}\binits{A.}~\bsnm{Agarwal}},
  \beditor{\bfnm{D.}\binits{D.}~\bsnm{Belgrave}},
  \beditor{\bfnm{K.}\binits{K.}~\bsnm{Cho}} \AND
  \beditor{\bfnm{A.}\binits{A.}~\bsnm{Oh}}, eds.).
\end{binproceedings}
\endbibitem

\bibitem[\protect\citeauthoryear{Iacus}{2008}]{iacus2008simulation}
\begin{bbook}[author]
\bauthor{\bsnm{Iacus},~\bfnm{Stefano~M}\binits{S.~M.}}
(\byear{2008}).
\btitle{Simulation and Inference for Stochastic Differential Equations: with
  {R} Examples}
\bvolume{486}.
\bpublisher{Springer}.
\end{bbook}
\endbibitem

\bibitem[\protect\citeauthoryear{Kappen, G{\'o}mez and
  Opper}{2012}]{kappen2012optimal}
\begin{barticle}[author]
\bauthor{\bsnm{Kappen},~\bfnm{Hilbert~J}\binits{H.~J.}},
  \bauthor{\bsnm{G{\'o}mez},~\bfnm{Vicen{\c{c}}}\binits{V.}} \AND
  \bauthor{\bsnm{Opper},~\bfnm{Manfred}\binits{M.}}
(\byear{2012}).
\btitle{Optimal control as a graphical model inference problem}.
\bjournal{Machine Learning}
\bvolume{87}
\bpages{159--182}.
\end{barticle}
\endbibitem

\bibitem[\protect\citeauthoryear{Kappen and Ruiz}{2016}]{kappen2016adaptive}
\begin{barticle}[author]
\bauthor{\bsnm{Kappen},~\bfnm{Hilbert~Johan}\binits{H.~J.}} \AND
  \bauthor{\bsnm{Ruiz},~\bfnm{Hans~Christian}\binits{H.~C.}}
(\byear{2016}).
\btitle{Adaptive importance sampling for control and inference}.
\bjournal{Journal of Statistical Physics}
\bvolume{162}
\bpages{1244--1266}.
\end{barticle}
\endbibitem

\bibitem[\protect\citeauthoryear{Karras et~al.}{2022}]{karraselucidating}
\begin{binproceedings}[author]
\bauthor{\bsnm{Karras},~\bfnm{Tero}\binits{T.}},
  \bauthor{\bsnm{Aittala},~\bfnm{Miika}\binits{M.}},
  \bauthor{\bsnm{Aila},~\bfnm{Timo}\binits{T.}} \AND
  \bauthor{\bsnm{Laine},~\bfnm{Samuli}\binits{S.}}
(\byear{2022}).
\btitle{Elucidating the Design Space of Diffusion-Based Generative Models}.
In \bbooktitle{Advances in Neural Information Processing Systems 35: Annual
  Conference on Neural Information Processing Systems 2022, NeurIPS 2022}
(\beditor{\bfnm{S.}\binits{S.}~\bsnm{Koyejo}},
  \beditor{\bfnm{S.}\binits{S.}~\bsnm{Mohamed}},
  \beditor{\bfnm{A.}\binits{A.}~\bsnm{Agarwal}},
  \beditor{\bfnm{D.}\binits{D.}~\bsnm{Belgrave}},
  \beditor{\bfnm{K.}\binits{K.}~\bsnm{Cho}} \AND
  \beditor{\bfnm{A.}\binits{A.}~\bsnm{Oh}}, eds.).
\end{binproceedings}
\endbibitem

\bibitem[\protect\citeauthoryear{Kessler, Lindner and
  S{\o}rensen}{2012}]{kessler2012statistical}
\begin{bbook}[author]
\bauthor{\bsnm{Kessler},~\bfnm{Mathieu}\binits{M.}},
  \bauthor{\bsnm{Lindner},~\bfnm{Alexander}\binits{A.}} \AND
  \bauthor{\bsnm{S{\o}rensen},~\bfnm{Michael}\binits{M.}}
(\byear{2012}).
\btitle{Statistical Methods for Stochastic Differential Equations}.
\bpublisher{Chapman \& Hall}.
\end{bbook}
\endbibitem

\bibitem[\protect\citeauthoryear{Kingma and Ba}{2014}]{kingma2014adam}
\begin{barticle}[author]
\bauthor{\bsnm{Kingma},~\bfnm{Diederik~P}\binits{D.~P.}} \AND
  \bauthor{\bsnm{Ba},~\bfnm{Jimmy}\binits{J.}}
(\byear{2014}).
\btitle{Adam: {A} method for stochastic optimization}.
\bjournal{arXiv preprint arXiv:1412.6980}.
\end{barticle}
\endbibitem

\bibitem[\protect\citeauthoryear{Klebaner}{2012}]{klebaner2012introduction}
\begin{bbook}[author]
\bauthor{\bsnm{Klebaner},~\bfnm{Fima~C}\binits{F.~C.}}
(\byear{2012}).
\btitle{Introduction to Stochastic Calculus with Applications}.
\bpublisher{World Scientific Publishing Company}.
\end{bbook}
\endbibitem

\bibitem[\protect\citeauthoryear{Kullback}{1968}]{kullback1968probability}
\begin{barticle}[author]
\bauthor{\bsnm{Kullback},~\bfnm{Solomon}\binits{S.}}
(\byear{1968}).
\btitle{Probability densities with given marginals}.
\bjournal{The Annals of Mathematical Statistics}
\bvolume{39}
\bpages{1236--1243}.
\end{barticle}
\endbibitem

\bibitem[\protect\citeauthoryear{L{\'e}ger}{2021}]{leger2021gradient}
\begin{barticle}[author]
\bauthor{\bsnm{L{\'e}ger},~\bfnm{Flavien}\binits{F.}}
(\byear{2021}).
\btitle{A gradient descent perspective on {S}inkhorn}.
\bjournal{Applied Mathematics \& Optimization}
\bvolume{84}
\bpages{1843--1855}.
\end{barticle}
\endbibitem

\bibitem[\protect\citeauthoryear{L{\'e}onard}{2012}]{leonard2012schrodinger}
\begin{barticle}[author]
\bauthor{\bsnm{L{\'e}onard},~\bfnm{Christian}\binits{C.}}
(\byear{2012}).
\btitle{From the {S}chr{\"o}dinger problem to the {M}onge--{K}antorovich
  problem}.
\bjournal{Journal of Functional Analysis}
\bvolume{262}
\bpages{1879--1920}.
\end{barticle}
\endbibitem

\bibitem[\protect\citeauthoryear{L{\'e}onard}{2014}]{leonard2014survey}
\begin{barticle}[author]
\bauthor{\bsnm{L{\'e}onard},~\bfnm{Christian}\binits{C.}}
(\byear{2014}).
\btitle{A survey of the {S}chr{\"o}dinger problem and some of its connections
  with optimal transport}.
\bjournal{Discrete and Continuous Dynamical Systems-Series A}
\bvolume{34}
\bpages{1533--1574}.
\end{barticle}
\endbibitem

\bibitem[\protect\citeauthoryear{Lipman et~al.}{2023}]{lipman2022flow}
\begin{binproceedings}[author]
\bauthor{\bsnm{Lipman},~\bfnm{Yaron}\binits{Y.}},
  \bauthor{\bsnm{Chen},~\bfnm{Ricky T.~Q.}\binits{R.~T.~Q.}},
  \bauthor{\bsnm{Ben{-}Hamu},~\bfnm{Heli}\binits{H.}},
  \bauthor{\bsnm{Nickel},~\bfnm{Maximilian}\binits{M.}} \AND
  \bauthor{\bsnm{Le},~\bfnm{Matthew}\binits{M.}}
(\byear{2023}).
\btitle{Flow Matching for Generative Modeling}.
In \bbooktitle{The Eleventh International Conference on Learning
  Representations, {ICLR} 2023, Kigali, Rwanda, May 1-5, 2023}.
\bpublisher{OpenReview.net}.
\end{binproceedings}
\endbibitem

\bibitem[\protect\citeauthoryear{Liu, Gong and Liu}{2023}]{liu2022flow}
\begin{binproceedings}[author]
\bauthor{\bsnm{Liu},~\bfnm{Xingchao}\binits{X.}},
  \bauthor{\bsnm{Gong},~\bfnm{Chengyue}\binits{C.}} \AND
  \bauthor{\bsnm{Liu},~\bfnm{Qiang}\binits{Q.}}
(\byear{2023}).
\btitle{Flow Straight and Fast: Learning to Generate and Transfer Data with
  Rectified Flow}.
In \bbooktitle{The Eleventh International Conference on Learning
  Representations, {ICLR} 2023, Kigali, Rwanda, May 1-5, 2023}.
\bpublisher{OpenReview.net}.
\end{binproceedings}
\endbibitem

\bibitem[\protect\citeauthoryear{Marin et~al.}{2012}]{marin2012approximate}
\begin{barticle}[author]
\bauthor{\bsnm{Marin},~\bfnm{Jean-Michel}\binits{J.-M.}},
  \bauthor{\bsnm{Pudlo},~\bfnm{Pierre}\binits{P.}},
  \bauthor{\bsnm{Robert},~\bfnm{Christian~P}\binits{C.~P.}} \AND
  \bauthor{\bsnm{Ryder},~\bfnm{Robin~J}\binits{R.~J.}}
(\byear{2012}).
\btitle{Approximate {B}ayesian computational methods}.
\bjournal{Statistics and Computing}
\bvolume{22}
\bpages{1167--1180}.
\end{barticle}
\endbibitem

\bibitem[\protect\citeauthoryear{Mikami}{2004}]{mikami2004monge}
\begin{barticle}[author]
\bauthor{\bsnm{Mikami},~\bfnm{Toshio}\binits{T.}}
(\byear{2004}).
\btitle{Monge's problem with a quadratic cost by the zero-noise limit of h-path
  processes}.
\bjournal{Probability Theory and Related Fields}
\bvolume{129}
\bpages{245--260}.
\end{barticle}
\endbibitem

\bibitem[\protect\citeauthoryear{{\O}ksendal}{2003}]{oksendal2003stochastic}
\begin{bbook}[author]
\bauthor{\bsnm{{\O}ksendal},~\bfnm{Bernt}\binits{B.}}
(\byear{2003}).
\btitle{Stochastic Differential Equations: An Introduction with Applications}.
\bpublisher{Springer}.
\end{bbook}
\endbibitem

\bibitem[\protect\citeauthoryear{Peluchetti}{2023}]{peluchetti2023diffusion}
\begin{barticle}[author]
\bauthor{\bsnm{Peluchetti},~\bfnm{Stefano}\binits{S.}}
(\byear{2023}).
\btitle{Diffusion Bridge Mixture Transports, Schr\"odinger Bridge Problems and
  Generative Modeling}.
\bjournal{arXiv preprint arXiv:2304.00917}.
\end{barticle}
\endbibitem

\bibitem[\protect\citeauthoryear{Roberts and
  Tweedie}{1996}]{roberts1996exponential}
\begin{barticle}[author]
\bauthor{\bsnm{Roberts},~\bfnm{Gareth~O}\binits{G.~O.}} \AND
  \bauthor{\bsnm{Tweedie},~\bfnm{Richard~L}\binits{R.~L.}}
(\byear{1996}).
\btitle{Exponential convergence of {L}angevin distributions and their discrete
  approximations}.
\bjournal{Bernoulli}
\bpages{341--363}.
\end{barticle}
\endbibitem

\bibitem[\protect\citeauthoryear{Rogers and
  Williams}{2000}]{rogers2000diffusions}
\begin{bbook}[author]
\bauthor{\bsnm{Rogers},~\bfnm{L~Chris~G}\binits{L.~C.~G.}} \AND
  \bauthor{\bsnm{Williams},~\bfnm{David}\binits{D.}}
(\byear{2000}).
\btitle{Diffusions, {M}arkov {P}rocesses and {M}artingales: {V}olume 2,
  {I}t{\^o} calculus}
\bvolume{2}.
\bpublisher{Cambridge university press}.
\end{bbook}
\endbibitem

\bibitem[\protect\citeauthoryear{R{\"u}schendorf}{1995}]{ruschendorf1995convergence}
\begin{barticle}[author]
\bauthor{\bsnm{R{\"u}schendorf},~\bfnm{Ludger}\binits{L.}}
(\byear{1995}).
\btitle{Convergence of the iterative proportional fitting procedure}.
\bjournal{The Annals of Statistics}
\bpages{1160--1174}.
\end{barticle}
\endbibitem

\bibitem[\protect\citeauthoryear{Salimans and
  Ho}{2022}]{salimans2022progressive}
\begin{binproceedings}[author]
\bauthor{\bsnm{Salimans},~\bfnm{Tim}\binits{T.}} \AND
  \bauthor{\bsnm{Ho},~\bfnm{Jonathan}\binits{J.}}
(\byear{2022}).
\btitle{Progressive Distillation for Fast Sampling of Diffusion Models}.
In \bbooktitle{The Tenth International Conference on Learning Representations,
  {ICLR} 2022, Virtual Event, April 25-29, 2022}.
\bpublisher{OpenReview.net}.
\end{binproceedings}
\endbibitem

\bibitem[\protect\citeauthoryear{Sharrock
  et~al.}{2022}]{sharrock2022sequential}
\begin{barticle}[author]
\bauthor{\bsnm{Sharrock},~\bfnm{Louis}\binits{L.}},
  \bauthor{\bsnm{Simons},~\bfnm{Jack}\binits{J.}},
  \bauthor{\bsnm{Liu},~\bfnm{Song}\binits{S.}} \AND
  \bauthor{\bsnm{Beaumont},~\bfnm{Mark}\binits{M.}}
(\byear{2022}).
\btitle{Sequential neural score estimation: Likelihood-free inference with
  conditional score based diffusion models}.
\bjournal{arXiv preprint arXiv:2210.04872}.
\end{barticle}
\endbibitem

\bibitem[\protect\citeauthoryear{Shi et~al.}{2022}]{shi2022conditional}
\begin{binproceedings}[author]
\bauthor{\bsnm{Shi},~\bfnm{Yuyang}\binits{Y.}},
  \bauthor{\bsnm{Bortoli},~\bfnm{Valentin~De}\binits{V.~D.}},
  \bauthor{\bsnm{Deligiannidis},~\bfnm{George}\binits{G.}} \AND
  \bauthor{\bsnm{Doucet},~\bfnm{Arnaud}\binits{A.}}
(\byear{2022}).
\btitle{Conditional simulation using diffusion Schr{\"{o}}dinger bridges}.
In \bbooktitle{Uncertainty in Artificial Intelligence, Proceedings of the
  Thirty-Eighth Conference on Uncertainty in Artificial Intelligence, {UAI}
  2022, 1-5 August 2022, Eindhoven, The Netherlands}
(\beditor{\bfnm{James}\binits{J.}~\bsnm{Cussens}} \AND
  \beditor{\bfnm{Kun}\binits{K.}~\bsnm{Zhang}}, eds.).
\bseries{Proceedings of Machine Learning Research}
\bvolume{180}
\bpages{1792--1802}.
\bpublisher{{PMLR}}.
\end{binproceedings}
\endbibitem

\bibitem[\protect\citeauthoryear{Shi et~al.}{2023}]{shi2023diffusion}
\begin{barticle}[author]
\bauthor{\bsnm{Shi},~\bfnm{Yuyang}\binits{Y.}},
  \bauthor{\bsnm{De~Bortoli},~\bfnm{Valentin}\binits{V.}},
  \bauthor{\bsnm{Campbell},~\bfnm{Andrew}\binits{A.}} \AND
  \bauthor{\bsnm{Doucet},~\bfnm{Arnaud}\binits{A.}}
(\byear{2023}).
\btitle{Diffusion Schr\"odinger Bridge Matching}.
\bjournal{arXiv preprint arXiv:2303.16852}.
\end{barticle}
\endbibitem

\bibitem[\protect\citeauthoryear{Sisson, Fan and
  Beaumont}{2018}]{sisson2018handbook}
\begin{bbook}[author]
\bauthor{\bsnm{Sisson},~\bfnm{Scott~A}\binits{S.~A.}},
  \bauthor{\bsnm{Fan},~\bfnm{Yanan}\binits{Y.}} \AND
  \bauthor{\bsnm{Beaumont},~\bfnm{Mark}\binits{M.}}
(\byear{2018}).
\btitle{Handbook of Approximate Bayesian Computation}.
\bpublisher{CRC Press}.
\end{bbook}
\endbibitem

\bibitem[\protect\citeauthoryear{Sohl{-}Dickstein
  et~al.}{2015}]{Sohl-DicksteinW15}
\begin{binproceedings}[author]
\bauthor{\bsnm{Sohl{-}Dickstein},~\bfnm{Jascha}\binits{J.}},
  \bauthor{\bsnm{Weiss},~\bfnm{Eric~A.}\binits{E.~A.}},
  \bauthor{\bsnm{Maheswaranathan},~\bfnm{Niru}\binits{N.}} \AND
  \bauthor{\bsnm{Ganguli},~\bfnm{Surya}\binits{S.}}
(\byear{2015}).
\btitle{Deep Unsupervised Learning using Nonequilibrium Thermodynamics}.
In \bbooktitle{Proceedings of the 32nd International Conference on Machine
  Learning, {ICML} 2015, Lille, France, 6-11 July 2015}
(\beditor{\bfnm{Francis~R.}\binits{F.~R.}~\bsnm{Bach}} \AND
  \beditor{\bfnm{David~M.}\binits{D.~M.}~\bsnm{Blei}}, eds.).
\bseries{{JMLR} Workshop and Conference Proceedings}
\bvolume{37}
\bpages{2256--2265}.
\bpublisher{JMLR.org}.
\end{binproceedings}
\endbibitem

\bibitem[\protect\citeauthoryear{Song et~al.}{2021a}]{song2020score}
\begin{binproceedings}[author]
\bauthor{\bsnm{Song},~\bfnm{Yang}\binits{Y.}},
  \bauthor{\bsnm{Sohl{-}Dickstein},~\bfnm{Jascha}\binits{J.}},
  \bauthor{\bsnm{Kingma},~\bfnm{Diederik~P.}\binits{D.~P.}},
  \bauthor{\bsnm{Kumar},~\bfnm{Abhishek}\binits{A.}},
  \bauthor{\bsnm{Ermon},~\bfnm{Stefano}\binits{S.}} \AND
  \bauthor{\bsnm{Poole},~\bfnm{Ben}\binits{B.}}
(\byear{2021}a).
\btitle{Score-Based Generative Modeling through Stochastic Differential
  Equations}.
In \bbooktitle{9th International Conference on Learning Representations, {ICLR}
  2021, Virtual Event, Austria, May 3-7, 2021}.
\bpublisher{OpenReview.net}.
\end{binproceedings}
\endbibitem

\bibitem[\protect\citeauthoryear{Song et~al.}{2021b}]{song2021maximum}
\begin{binproceedings}[author]
\bauthor{\bsnm{Song},~\bfnm{Yang}\binits{Y.}},
  \bauthor{\bsnm{Durkan},~\bfnm{Conor}\binits{C.}},
  \bauthor{\bsnm{Murray},~\bfnm{Iain}\binits{I.}} \AND
  \bauthor{\bsnm{Ermon},~\bfnm{Stefano}\binits{S.}}
(\byear{2021}b).
\btitle{Maximum Likelihood Training of Score-Based Diffusion Models}.
In \bbooktitle{Advances in Neural Information Processing Systems 34: Annual
  Conference on Neural Information Processing Systems 2021, NeurIPS 2021,
  December 6-14, 2021, virtual}
(\beditor{\bfnm{Marc'Aurelio}\binits{M.}~\bsnm{Ranzato}},
  \beditor{\bfnm{Alina}\binits{A.}~\bsnm{Beygelzimer}},
  \beditor{\bfnm{Yann~N.}\binits{Y.~N.}~\bsnm{Dauphin}},
  \beditor{\bfnm{Percy}\binits{P.}~\bsnm{Liang}} \AND
  \beditor{\bfnm{Jennifer~Wortman}\binits{J.~W.}~\bsnm{Vaughan}}, eds.)
\bpages{1415--1428}.
\end{binproceedings}
\endbibitem

\bibitem[\protect\citeauthoryear{Song et~al.}{2023}]{pmlr-v202-song23k}
\begin{binproceedings}[author]
\bauthor{\bsnm{Song},~\bfnm{Jiaming}\binits{J.}},
  \bauthor{\bsnm{Zhang},~\bfnm{Qinsheng}\binits{Q.}},
  \bauthor{\bsnm{Yin},~\bfnm{Hongxu}\binits{H.}},
  \bauthor{\bsnm{Mardani},~\bfnm{Morteza}\binits{M.}},
  \bauthor{\bsnm{Liu},~\bfnm{Ming-Yu}\binits{M.-Y.}},
  \bauthor{\bsnm{Kautz},~\bfnm{Jan}\binits{J.}},
  \bauthor{\bsnm{Chen},~\bfnm{Yongxin}\binits{Y.}} \AND
  \bauthor{\bsnm{Vahdat},~\bfnm{Arash}\binits{A.}}
(\byear{2023}).
\btitle{Loss-Guided Diffusion Models for Plug-and-Play Controllable
  Generation}.
In \bbooktitle{Proceedings of the 40th International Conference on Machine
  Learning}
(\beditor{\bfnm{Andreas}\binits{A.}~\bsnm{Krause}},
  \beditor{\bfnm{Emma}\binits{E.}~\bsnm{Brunskill}},
  \beditor{\bfnm{Kyunghyun}\binits{K.}~\bsnm{Cho}},
  \beditor{\bfnm{Barbara}\binits{B.}~\bsnm{Engelhardt}},
  \beditor{\bfnm{Sivan}\binits{S.}~\bsnm{Sabato}} \AND
  \beditor{\bfnm{Jonathan}\binits{J.}~\bsnm{Scarlett}}, eds.).
\bseries{Proceedings of Machine Learning Research}
\bvolume{202}
\bpages{32483--32498}.
\bpublisher{PMLR}.
\end{binproceedings}
\endbibitem

\bibitem[\protect\citeauthoryear{Tzen and Raginsky}{2019}]{tzen2019theoretical}
\begin{binproceedings}[author]
\bauthor{\bsnm{Tzen},~\bfnm{Belinda}\binits{B.}} \AND
  \bauthor{\bsnm{Raginsky},~\bfnm{Maxim}\binits{M.}}
(\byear{2019}).
\btitle{Theoretical guarantees for sampling and inference in generative models
  with latent diffusions}.
In \bbooktitle{Conference on Learning Theory, {COLT} 2019, 25-28 June 2019,
  Phoenix, AZ, {USA}}
(\beditor{\bfnm{Alina}\binits{A.}~\bsnm{Beygelzimer}} \AND
  \beditor{\bfnm{Daniel}\binits{D.}~\bsnm{Hsu}}, eds.).
\bseries{Proceedings of Machine Learning Research}
\bvolume{99}
\bpages{3084--3114}.
\bpublisher{{PMLR}}.
\end{binproceedings}
\endbibitem

\bibitem[\protect\citeauthoryear{Vargas, Grathwohl and
  Doucet}{2023}]{vargasDDSampler2023}
\begin{binproceedings}[author]
\bauthor{\bsnm{Vargas},~\bfnm{Francisco}\binits{F.}},
  \bauthor{\bsnm{Grathwohl},~\bfnm{Will~Sussman}\binits{W.~S.}} \AND
  \bauthor{\bsnm{Doucet},~\bfnm{Arnaud}\binits{A.}}
(\byear{2023}).
\btitle{Denoising Diffusion Samplers}.
In \bbooktitle{The Eleventh International Conference on Learning
  Representations, {ICLR} 2023, Kigali, Rwanda, May 1-5, 2023}.
\bpublisher{OpenReview.net}.
\end{binproceedings}
\endbibitem

\bibitem[\protect\citeauthoryear{Vargas et~al.}{2021}]{vargas2021solving}
\begin{barticle}[author]
\bauthor{\bsnm{Vargas},~\bfnm{Francisco}\binits{F.}},
  \bauthor{\bsnm{Thodoroff},~\bfnm{Pierre}\binits{P.}},
  \bauthor{\bsnm{Lamacraft},~\bfnm{Austen}\binits{A.}} \AND
  \bauthor{\bsnm{Lawrence},~\bfnm{Neil}\binits{N.}}
(\byear{2021}).
\btitle{Solving {S}chr{\"o}dinger bridges via maximum likelihood}.
\bjournal{Entropy}
\bvolume{23}
\bpages{1134}.
\end{barticle}
\endbibitem

\bibitem[\protect\citeauthoryear{Vargas et~al.}{2023}]{vargas2023bayesian}
\begin{barticle}[author]
\bauthor{\bsnm{Vargas},~\bfnm{Francisco}\binits{F.}},
  \bauthor{\bsnm{Ovsianas},~\bfnm{Andrius}\binits{A.}},
  \bauthor{\bsnm{Fernandes},~\bfnm{David}\binits{D.}},
  \bauthor{\bsnm{Girolami},~\bfnm{Mark}\binits{M.}},
  \bauthor{\bsnm{Lawrence},~\bfnm{Neil~D}\binits{N.~D.}} \AND
  \bauthor{\bsnm{N{\"u}sken},~\bfnm{Nikolas}\binits{N.}}
(\byear{2023}).
\btitle{Bayesian learning via neural {S}chr{\"o}dinger--{F}{\"o}llmer flows}.
\bjournal{Statistics and Computing}
\bvolume{33}
\bpages{3}.
\end{barticle}
\endbibitem

\bibitem[\protect\citeauthoryear{Vincent}{2011}]{vincent2011connection}
\begin{barticle}[author]
\bauthor{\bsnm{Vincent},~\bfnm{Pascal}\binits{P.}}
(\byear{2011}).
\btitle{A connection between score matching and denoising autoencoders}.
\bjournal{Neural Computation}
\bvolume{23}
\bpages{1661--1674}.
\end{barticle}
\endbibitem

\bibitem[\protect\citeauthoryear{Zhang and
  Chen}{2022}]{zhangyongxinchen2021path}
\begin{binproceedings}[author]
\bauthor{\bsnm{Zhang},~\bfnm{Qinsheng}\binits{Q.}} \AND
  \bauthor{\bsnm{Chen},~\bfnm{Yongxin}\binits{Y.}}
(\byear{2022}).
\btitle{Path Integral Sampler: {A} Stochastic Control Approach For Sampling}.
In \bbooktitle{The Tenth International Conference on Learning Representations,
  {ICLR} 2022, Virtual Event, April 25-29, 2022}.
\bpublisher{OpenReview.net}.
\end{binproceedings}
\endbibitem

\bibitem[\protect\citeauthoryear{Zhang, Sahai and
  Marzouk}{2021}]{zhangmarzouk2021sampling}
\begin{binproceedings}[author]
\bauthor{\bsnm{Zhang},~\bfnm{Benjamin}\binits{B.}},
  \bauthor{\bsnm{Sahai},~\bfnm{Tuhin}\binits{T.}} \AND
  \bauthor{\bsnm{Marzouk},~\bfnm{Youssef}\binits{Y.}}
(\byear{2021}).
\btitle{Sampling via controlled stochastic dynamical systems}.
In \bbooktitle{I (Still) Can't Believe It's Not Better! NeurIPS 2021 Workshop}.
\end{binproceedings}
\endbibitem

\end{thebibliography}


\end{document}